\documentclass[final,5p,times,twocolumn,authoryear,runningheads]{elsarticle}
\usepackage{graphicx}
\usepackage{txfonts}
\usepackage{color}
\usepackage{natbib}
\usepackage[switch, modulo]{lineno}
\usepackage{rotating}
\usepackage{xspace}
\usepackage{float}
\usepackage{url}
\usepackage{multirow}
\usepackage{gensymb}
\usepackage{amssymb}
\usepackage{amsmath}
\usepackage{threeparttable}
\newcommand{\g}{$\gamma$\xspace}

\begin{document}

\begin{frontmatter}
\title{Prospects on high-energy source searches based on pattern recognition:\\ \vspace{0.2cm}\normalsize Object detection in the H.E.S.S. Galactic Plane Survey and catalogue cross-matches}

\author[1,2]{Q.~Remy\corref{cor1}}%
\ead{quentin.remy@mpi-hd.mpg.de}
\author[1]{Y.~A.~Gallant}
\author[1]{M.~Renaud}
\cortext[cor1]{Corresponding author}
\address[1]{~Laboratoire Univers et Particules de Montpellier, CNRS/IN2P3, Universit\'e de Montpellier, F-34095 Montpellier, France}
\address[2]{~Max-Planck-Institut f\"ur Kernphysik, P.O. Box 103980, D 69029 Heidelberg, Germany}
\date{\today}
 
\begin{abstract}
%\textit{Context.}
The H.E.S.S. Galactic Plane Survey \citep[HGPS,][]{2018A&A...612A...1H} represents one of the most sensitive surveys of the Galactic Plane at very high energies (VHE, 0.1 $<$ E $<$ 100 TeV). However the source detection algorithm of the HGPS pipeline is not well-suited for complex regions, including sources with shell-like morphologies. As an alternative and complementary approach, we have investigated blind search methods for VHE $\gamma$-ray source detection based on well-known and widely used image processing and pattern recognition techniques. %\\

%\textit{Aims.}

Our goal is to build in a short amount of computational time a list of potentially valuable objects without prior case-specific morphological assumptions. We aim to classify and rank the detected objects in order to identify only the most promising source candidates for further multi-wavelength-association searches, dedicated analyses, or deeper observations. %\\
 
%\textit{Methods.}
In the approach proposed, we extract sparse and pertinent structural information from the significance maps using a edge detection operator. We then apply a Hough circle transform and detect a collection of objects as local maxima in the Hough space. On the basis of morphological parameters we can characterize different object classes. Classification can be used to identify valuable source candidates sharing the characteristics of well-known sources. %\\

%\textit{Results.} 
We show that using these pattern recognition techniques we can detect objects with partial circular symmetry irrespective of a morphological template (e.g.\ point-like, Gaussian-like, or shell-like). All the shell-type supernova remnants (SNRs) catalogued in the HGPS (from dedicated analyses) are associated with at least one detected object. Catalogue cross-matches indicate that several detected objects not catalogued in the HGPS are spatially coincident with multi-wavelength counterparts. %\\

%\textit{Conclusions.} 
This paper can be seen as a prospective study for the search of VHE \g-ray sources based on Hough transform and morphological classification. The algorithm have been tested on bootstrap simulations and applied to significance maps of the H.E.S.S. Galactic survey. Further investigation on the most promising candidates will be conducted in dedicated follow-up analyses.
\end{abstract}

\begin{keyword}
Gamma rays: Sources \sep TeV: supernova remnants \sep Methods: data analysis
\end{keyword}
%\titlerunning{BLOBfind: blind search for TeV sources and catalogues cross-matches}
%\authorrunning{Remy et al.}

\end{frontmatter}

\section{Introduction}

The H.E.S.S. Galactic Plane Survey \citep[HGPS, ][]{2018A&A...612A...1H} represents the one of the most sensitive surveys of the Galactic plane (-110\degree~$<$~$l$~$<$~65\degree, $|b|$~$\leq$~3\degree) in the very high energy (VHE) \g-ray domain (0.1--100 TeV). The HGPS source catalogue includes 64 sources detected with the HGPS analysis pipeline and 14 additional sources analysed independently. The latter have been excluded from the HGPS source detection and analysis pipeline because of their complexity, and include in particular sources in the Galactic Centre region \citep{2005A&A...432L..25A,2016Natur.531..476H,2018A&A...612A...9H} and sources with shell-like morphologies \citep{2011A&A...531A..81H,2018A&A...612A...4H,2018A&A...612A...6H,2018A&A...612A...7H,2018A&A...612A...8H}.

The HGPS catalog differentiates four classes of identified sources: pulsar wind nebulae (PWNe), (shell-type) supernova remnants (SNRs), composite SNRs, and \g-ray binaries. The SNR class includes both well-resolved shell-type SNRs and unresolved VHE sources, or sources with unclear morphology, for which the association of the VHE emission with a SNR is highly probable (as in the case of W28, W49B and G353.6-0.7). PWNe refer to plerionic objects with no observed shell, or to resolved plerionic components in composite SNRs, which consist of a PWN inside an SNR shell.  HGPS sources classified as composites refer to VHE sources which are spatially coincident with a known composite SNR, but which are not sufficiently well-resolved to distinguish between the PWN and the SNR shell as the origin of the VHE emission \citep{2018A&A...612A...1H}.
% \citep{2018A&A...612A...2H,2018A&A...612A...8H}.

The source detection procedure applied to the H.E.S.S.\ GPS is based on iterative template fitting. Starting with an empirical model of the diffuse background emission, the most significant \g-ray excesses have been iteratively fitted with two-dimensional symmetric Gaussian components by means of a maximum-likelihood estimation \citep{2018A&A...612A...1H}. The last step of the HGPS catalogue pipeline consists in merging the overlapping Gaussian components, not clearly resolved as separate emission peaks, into single sources. Such an approach presents two major drawbacks: on the technical side, the convergence of the iterative scheme is time consuming, especially as the computation time greatly increases with the number of objects and the associated confusion; on the physics side, the morphological description of an extended $\gamma$-ray source is limited by the simplistic template in use. In particular, as Gaussian morphology is imposed, a search for shell-like source has been conducted independently \citep{2018A&A...612A...8H}. Such an approach limits the capability to detect complex-shape or nested sources, and could induce a selection bias in further population studies. 

In this paper we propose an alternative object detection approach, complementary to the iterative template-fitting method. 
We have investigated blind search methods for VHE $\gamma$-ray source detection based on well-known and widely used image processing and pattern recognition techniques. Our goal is to build in a short amount of computational time a list of potentially valuable objects without prior case-specific morphological assumptions. This list of objects could be used to:

\begin{itemize}
\item[-] provide seed source candidates with robust guess on their position and morphological parameters to be tested by conventional analysis pipelines and/or dedicated analyses using template-fitting approach \citep[as done for the \textit{Fermi}-LAT point-source catalogues][but generalized to extended sources]{2017ApJS..232...18A}; \vspace{0.1cm}

\item[-] give a detailed description of the sub-structures in complex regions that would ease multi-wavelength (MW) association search and so help firmly identify multiple nested sources; \vspace{0.1cm}

\item[-] evaluate potential valuable targets to support additional observation proposals, in particular toward source candidates below the significance threshold at the sensitivity of current observations that could be confirmed as sources by deeper surveys. \vspace{0.1cm}

\end{itemize}

In the following we show that Hough circle transform can be used to detect objects with partial circular symmetry regardless of a shell-like or Gaussian-like morphology. We have tested the robustness and the efficiency of the algorithm to reconstruct the position and extent of simulated sources and to detect the firmly identified sources in the HGPS catalogue. We also investigate how the morphological parameters of the detected objects can be used to differentiate the various source populations in the HGPS. We discuss on the identification of new source candidates based on such morphological classification, and/or supported by spatial coincidence search in MW catalogues.

The paper is structured as follows: In Sect.~\ref{sec:data} we present the data and catalogues used; In Sect.~\ref{sec:MthDet} we describe the detection algorithm based on Hough circle transform, the parameters derived in order to classify the detected objects, and the association procedure. In Sect.~\ref{sec:SimTest} we discuss the tests on bootstrap simulations. In Sect.~\ref{sec:HGPScomp}, we present the list of objects found in the HGPS maps compared to the catalogued sources; In Sect.~\ref{sec:discu}, we discuss the identification of the objects based on the morphological classification and MW coincidence search; In Sect.~\ref{sec:ccl} we briefly summarize the results and discuss possible future studies.
In supplements, the production of the simulated skies is detailed in \ref{apx:Sim_data}; potential improvements of the algorithm are discussed in \ref{apx:improvements}; and additional tables and figures are given in \ref{apx:add}. 

\section{Data}
\label{sec:data}

\subsection{H.E.S.S. Galactic Plane Survey (HGPS) maps} 

We have used significance and flux maps of the HGPS survey\footnote{available online at: \url{https://www.mpi-hd.mpg.de/hfm/HESS/hgps/}} presented in \cite{2018A&A...612A...1H}. Pixels of the significance maps contain the values of the statistical significance of the \g-ray excess in the sense of \cite{1983ApJ...272..317L} within a given correlation radius R$_{corr}$. The background level is estimated by counting the number of photons in OFF-regions, bearing in mind that the large-scale emission from the Galaxy is not taken into account in this computation. The value of a given pixel in the flux maps corresponds to the integrated flux above 1 TeV of a potential source centered on the pixel and fully enclosed within the correlation radius of the map, assuming a power-law spectrum with a spectral index of 2.3 \citep[see Eq. 2 to 4 in][for more details]{2018A&A...612A...1H}.

\subsection{Catalogues}
\label{sec:cats}

In order to associate the objects detected by our algorithm to known sources at various wavelengths, we have made use of the following catalogues:

\begin{itemize}

\item[-] The HGPS source catalogue, containing 78 sources of VHE \g-rays \citep{2018A&A...612A...1H}. The majority of the HGPS sources are spatially coincident with known sources at other wavelengths that could potentially account for the production of \g-rays at VHE energies but only 40\% of the HGPS sources could be firmly identified. Among the 31 firmly identified sources there are 12 PWNe, 8 SNRs, 8 composite SNRs (where both the interior PWN and SNR shell may contribute to the emission) and 3 high-energy binary systems. Most of the so-called unidentified sources are associated with multiple objects but one cannot pinpoint the precise origin of the VHE emission because of the source confusion induced by the relatively broad point-spread function (PSF) of VHE \g-ray observations compared to those in other domains. The unassociated sources could emit exclusively in the VHE domain (truly dark sources) or may have been missed at other (radio, X-ray) wavelengths. If so, they would require either a more sophisticated MW association procedure or deeper observations.\vspace{0.1cm}

\item[-] the gamma-cat package\footnote{\url{http://gamma-cat.readthedocs.io/}} (version 0.1, July 2018) which provides a collection of data and catalogues for VHE \g-ray astronomy. \vspace{0.1cm}.
\item[-] the SNRcat\footnote{\url{http://snrcat.physics.umanitoba.ca/}}, including Galactic SNRs and PWNe \citep{2012AdSpR..49.1313F}. The list of SNRs with their basic physical properties is based on the Catalogue of Galactic SNRs by \cite{2014BASI...42...47G} and on the list of Galactic SNRs Interacting with Molecular Clouds maintained by Bing Jiang\footnote{\url{http://astronomy.nju.edu.cn/~ygchen/others/bjiang/interSNR6.htm}}. Entries have also been cross-checked with the Pulsar Wind Nebula catalogue \citep{2006csxs.book..279K} and the SGR/AXP catalogue from the McGill Pulsar Group \citep{2014ApJS..212....6O}. This catalogue is updated regularly, we used the same version as that considered in the HGPS catalogue (from Oct 10, 2015). \vspace{0.1cm}

\item[-] the \textit{Fermi}-LAT eight-year source catalog\footnote{\url{https://fermi.gsfc.nasa.gov/ssc/data/access/lat/8yr_catalog/}} \citep[4FGL v19, ][]{2019arXiv190210045T} based on eight years of data in the 100 MeV -- 1 TeV range. The 4FGL catalog includes 5065 sources above 4$\sigma$ significance. Seventy-five sources are modelled explicitly as spatially extended. Compared to the 3FGL catalogue \citep{2015ApJS..218...23A}, the 4FGL source list has twice as much exposure as well as a number of analysis improvements, including an updated model for the Galactic diffuse \g-ray emission\footnote{\url{https://fermi.gsfc.nasa.gov/ssc/data/analysis/software/aux/4fgl/Galactic_Diffuse_Emission_Model_for_the_4FGL_Catalog_Analysis.pdf}} .\vspace{0.1cm}

\item[-] the first \textit{Fermi}-LAT Supernova Remnant Catalogue\footnote{\url{https://fermi.gsfc.nasa.gov/ssc/data/access/lat/1st_SNR_catalog/}} (SC1). \cite{2016ApJS..224....8A} have analysed 36 months of \textit{Fermi}-LAT \g-ray data in the 1--100 GeV range toward 279 regions containing known radio SNRs. They found 102 source candidates among which 30 have sufficient spatial overlap with their corresponding radio SNRs and significance with respect to various diffuse emission models to suggest these are the GeV counterparts, and 14 additional candidates which may also be related to known radio-emitting SNRs.\vspace{0.1cm}

\end{itemize}

\section{Methods}
\label{sec:MthDet}

\subsection{Object detection in Hough space}

We have applied the object detection algorithm to significance maps with R$_{corr}$ = 0.1\degree~from the HGPS survey and bootstrap simulations. We have performed a single scan of the large region of the sky covered by the HGPS in order to detect potential VHE \g-ray sources in a uniform and automated way. In the following sub-sections we detail the algorithm step by step.

\subsubsection{Pre-processing}
\label{sec:prepro}

The goal of the pre-processing is to extract a sparse description of the relevant structures in the image that can be used to compute the Hough transform efficiently. In practice it consists in applying an edge detection operator, such as the Canny filter \citep{Canny1986ACA}.

\begin{figure*}
\centering 
\includegraphics[width=\hsize]{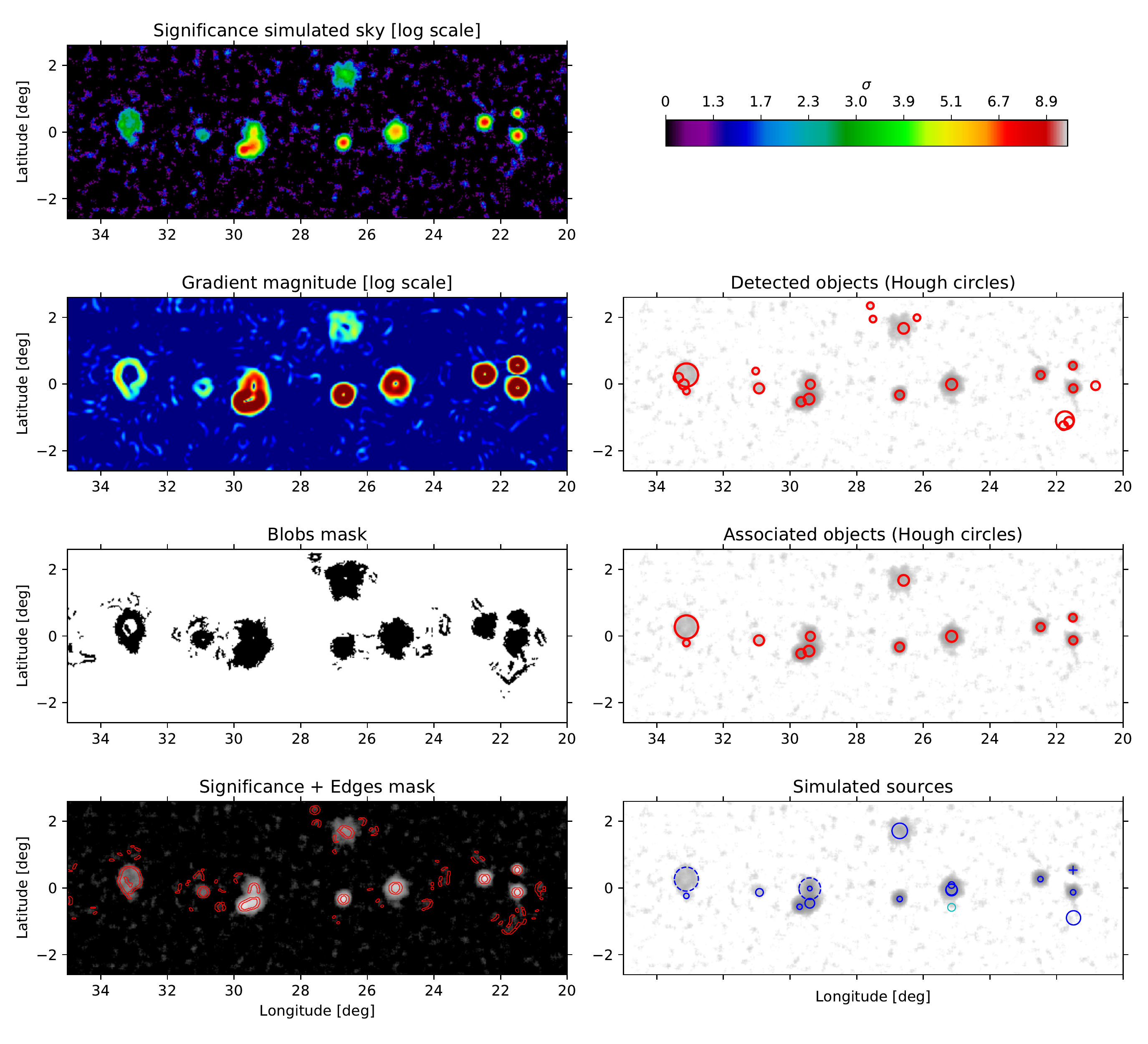}  
\caption{
Objects detection and association -- These figures illustrate the application of the method to a simulated sky (see Sect. \ref{sec:SimTest} and appendix \ref{apx:Sim_data}). First row, left: Significance map derived using the Adaptive Ring Background method as in the HGPS (R$_{corr}$ = 0.1\degree). Second row, left: Gradient of the significance map derived using Sobel operator. Third row, left: Blob mask given by thresholding the significance and gradient values (as described in Sect.~\ref{sec:prepro}). Fourth row, left: Significance map at R$_{corr}$ = 0.1\degree~(same as first row but in grey-scale) with the Edge mask obtained after applying the Canny filter overlaid in red. Second row, right: objects detected in Hough space. Third row, right: Detected objects associated to simulated sources  using criterion on inter-center distance and surface overlap as described in Sect.~\ref{sec:xmatch}. Note that we test for associations regardless of the expected TS values. Fourth row, right : Simulated sources, the dotted circles correspond to the outer radii of radial shells, the plain circles correspond to the sigma of radial Gaussians, point sources are shown as crosses. Light and dark blue colors correspond to sources with an expected TS value below and above 30, respectively.
}
\label{fig:mth_ex}
\end{figure*}

The edge detection procedure is described as follows: First we smooth the significance image by applying a convolution with a Gaussian kernel ($\sigma = 0.09\degree=4.5$ pixels). Then we use the Sobel-Feldman operator\footnote{\url{https://www.researchgate.net/publication/239398674_An_Isotropic_3_3_Image_Gradient_Operator}} \citep{Sobel1968} to compute an approximation of the image gradient. For an image I, the Sobel-Feldman operator returns a value for the first derivative in the horizontal direction ($G_x$) and the vertical direction ($G_y$) as :
\begin{equation}
G_x= I \ast \begin{bmatrix}
   -1 & 0 & 1 \\
   -2 & 0 & 2  \\
   -1 & 0 & 1 
   \end{bmatrix}
,    \quad
G_y= I \ast \begin{bmatrix}
   -1 & -2 & -1 \\
    0 & 0  & 0 \\
    1 & 2 & 1 
   \end{bmatrix}   
\end{equation}
At each point in the image, the gradient magnitude is then derived as $G= \sqrt{G_x^2+G_y^2}$, and the gradient direction as $\theta=\arctan(G_y/G_x)$. The gradient magnitude provide thick edges of the structures so a non-maximum suppression is applied to obtain 1-pixel thin edges. These first steps are performed following the implementation of the Canny filter in scikit-image python package\footnote{for further details see \url{https://scikit-image.org/docs/dev/api/skimage.feature.html?highlight=feature\%20canny#skimage.feature.canny}} while the subsequent steps differ. Then we filter the thin edges using conditions on both gradient and significance values. We mask the thin edges where gradient magnitude is lower than its 86.6 percentile value ($\sim1.5\sigma$ for a Gaussian distribution). The significance maps are filtered by hysteresis thresholding with a high threshold of 4.5$\sigma$  and a low threshold of 1$\sigma$ (pixels between these two thresholds are masked only if they are not connected to a pixel above the high threshold).
Contrary to the hard threshold usually applied in VHE \g-ray studies for source detection (as in the HGPS procedure), a hysteresis thresholding better preserves the global spatial coherence of the extended excess even if the whole structure is not strictly above the hard threshold. In order to exploit the information of the significance maps with R$_{corr}$ = 0.1\degree~and R$_{corr}$ = 0.2\degree~we combine their respective hysteresis masks with an \textit{or} condition.  The boolean mask (named "blob mask") associating the gradient and significance filters is shown in Fig \ref{fig:mth_ex} and \ref{fig:Sim_preprocess} (third panel) for one of the simulated maps.
Finally, we removed the small edges of less than 3 pixels long. The final result is a boolean mask (named "edge mask") highlighting the sharp edges of the significance map, as illustrated in Fig. \ref{fig:mth_ex} and \ref{fig:Sim_preprocess}.

The structure of the gradient associated to the diffuse background has a latitudinal symmetry centred on the Galactic Plane, while the sources can be seen as spherical perturbations to this background gradient. In order to improve the contrast between the source and the diffuse structures we applied a longitudinal thickening to the edge mask: for each pixel in the edge mask we extend the mask by one pixel (i.e. 0.02\degree) along the longitude axis, so that the sources appear on average thicker than the diffuse structures, which often already appears as longitudinal features. This simple step reduces the weight of diffuse structure compared to sources in Hough space, and should reduce the false detections triggered by the diffuse background (particularly toward the dense clouds of the Central Molecular Zone in the vicinity of the Galactic Center, see second line of Fig. \ref{fig:Comparisons_HESS}).
%Using the gradient information to extract only sparse structural information and detect only sharp edges in the data allows in principle one to filter the diffuse structures associated to the Galactic background. The fact that the algorithm does not trigger detection all along the VHE-bright and diffuse Central Molecular Zone supports this idea.

\subsubsection{Finding local maxima in Hough space}
\label{sec:locmax}

The classical Hough transform is meant to identify lines in an image, and was introduced by \cite{Hough:1959qva} to detect particle trajectories in bubble chamber images. The generalized Hough transform proposed by \cite{Duda1972UHT} extends the principle to any analytically defined shapes, usually circles or ellipses. More recent implementations generalize the concept to arbitrary complex shapes by performing template matching \citep{BALLARD81}. For our purpose a simple Hough circle transform can be used to indifferently detect sources with point-like, Gaussian-like or shell-like morphologies.

For a circle transform the Hough space is represented by a 3D accumulator matrix in longitude, latitude, radius. The accumulator resolution is set to 0.04\degree~in longitude-latitude and 0.01\degree~in radius. We fix the minimal radius to 0.1\degree, to match the minimal correlation radius of the significance maps used. These choices limit the computational time to less than 5 minutes to process the whole HGPS region, without limiting the sampling precision which remains lower than the H.E.S.S. angular resolution.

In practice for each point of the edge mask, $(l_i,b_i)$, and for each radius sampled, $R_i$, we can define a circle, $C(l_i,b_i,R_i)$, in the Hough space. So the values of the accumulator matrix, HA$(l,b,R)$, (initially zeros) are incremented toward the pixels where the circle passes through: $HA(l_C,b_C,R_i)$+=1 $\forall (l_C,b_C) \in C$. The angle sampling used to compute the circles is limited to 1\degree. We normalize the accumulator values by the angle sampling dimension, so the value of the accumulator can be assimilated to a fraction of the circle circumference in the original space. In the following, we will refer to this parameter as the circularity of the object. Once the accumulator matrix is built the next step is to find its local maximum. To do so we select only the pixels that are at maximum within a 3D sliding window of $0.3\degree\times0.3\degree\times0.5\degree$ in longitude-latitude-radius or $0.5\degree\times0.5\degree\times0.5\degree$ if their radius is larger than 0.25\degree~(we use two window sizes to scan different structure scales). We discard objects with circularity lower than 0.6 or 0.5 if their radius is larger than 0.25\degree~(so the detected objects are at least half-circular). We also discard objects with radius lower than 0.25\degree~if the significance at their center is not larger than 3$\sigma$ in significance maps with R$_{corr}$ = 0.1\degree~or R$_{corr}$ = 0.2\degree. Finally we remove potential duplicates with both inter-center distance and radius difference lower than 0.1 \degree. The position of a local maximum in the Hough space gives the centre and the radius of the object in the original space. 

The different threshold values introduced here and previously in Sect. \ref{sec:prepro} have been empirically optimized in order to maximize the fraction of the simulated sources which are well-reconstructed, given the association criterion introduced in Sect~\ref{sec:xmatch}. We have also ensured to maintain the fraction of detected objects associated to a simulated source higher than about 30\% in average (or similarly the detected objects to significant sources ratio lower than about three, see discussion in Sect. \ref{sec:discl}).  

The resulting list of objects provides a discrete and parametrised view of the pertinent structures in the data that is easier to interpret and analyse than the continuous values of a significance map. However the detected objects cannot be directly interpreted as sources. In order to identify the most valuable candidates, that could warrant a dedicated analysis, several approaches are possible: distinguish the sub-structures from the main objects, guess the source type by comparing its morphological parameters to well-identified sources, and find possible associations at other wavelengths. We further discuss these issues in Sect.~\ref{sec:discu}.

\subsection{Classification based on morphological parameters}
\label{sec:params}

Determining a class of objects that can be assimilated to a known source population is a well-suited problem for machine-learning techniques. Classification requires a parameter space in which the different populations can be isolated.
We introduce the following set of morphological parameters:

\begin{itemize}

\item[-] the radius given by the third coordinate in the Hough accumulator space;
\item[-] significance at central pixel of the object (in the map with R$_{corr}$ = 0.1$\degree$);
\item[-] the circularity given by the Hough accumulator value;
\item[-] the Pearson correlation coefficient ($PCC$ or Pearson-R) of a 5-point radial profile in flux: for each object we have integrated the flux map (with R$_{corr}$ = 0.1$\degree$) in 5 rings of equal area between the center of the object and its radius;
\item[-] a morphology flag, the Pearson-R coefficient of the radial profile in flux is used to distinguish 3 types of structures:
\begin{itemize}
\item[] Peak $\:\, \equiv \: PCC\, <-1/3$,
\item[] Flat $\;\;\: \equiv |PCC|<1/3$,
\item[] Cavity $\equiv \: PCC\, >1/3$;
\end{itemize}
\item[-] a nesting flag: for each pair of objects $C$ of radius $R$ and $R_{sub}$, we calculate their inter-center distance, $d$, and define 5 overlapping configurations :
\begin{itemize}
\item[] Class 0 $\equiv C(R): d>R+R_{sub}, \; \forall \, C(R_{sub})$,
\item[] Class 1 $\equiv C(R): \exists \, C(R_{sub}) \,\, \,| \; d \leqslant R+R_{sub}, \; R_{sub} \leqslant R $, 
\item[] Class 2 $\equiv C(R_{sub})\,: \exists \, C(R) \, \,| \; d \leqslant R+R_{sub}, \; R_{sub}<R$,
\item[] Class 3 $\equiv C(R_{sub})\,: \exists \, C(R) \, \,| \; d \leqslant R, \; R_{sub}<R$,
\item[] Class 4 $\equiv C(R_{sub})\,: \exists \, C(R) \, \,| \; d \leqslant R-R_{sub}, \; R_{sub}<R $.
\end{itemize}

\end{itemize}

In the following, Class 0 and 1 are defined as main objects; Class 0 objects are non-overlapping, while Class 1 are large objects partially overlapping with smaller ones.  These smaller objects are considered as sub-structures, and are of Class 2, 3 or 4 depending on the inter-center distance to the main object.

From a basic point-of-view, the pertinence of a Gaussian-like object with R $<$ 0.2\degree~can be filtered by threshold conditions in significance at its center, the pertinence of a shell-like object can be filtered by threshold conditions in circularity and Pearson-R. The unresolved objects should reach minimal radius and maximal circularity. The potential of automated object classification and source population identification on this basis is further discussed in Sect.~\ref{sec:discu}. We will present simple classification schemes based on the comparison of detected objects with well-identified H.E.S.S.~sources in the parameter space introduced here.

\subsection{Catalogue cross-matches}
\label{sec:xmatch}

In order to associate the detected objects with the known sources we test for spatial coincidence using two criteria based on inter-center distance and surface overlap. For each object we search for known sources within an inter-distance:
\mbox{$d_{c}<0.1+0.3\times R_{object}$}
and we report only the source maximizing the surface overlap fraction, defined as:
\begin{equation}
\rm{SF_{overlap}}=\frac{S_{object} \cap S_{source}}{S_{object} \cup S_{source}}.
\end{equation}
We choose to report only the strongest positional coincidence in order to limit the possible associations for extended objects. As the objects are detected on significance maps with a correlation radius of $R_{corr}=0.1 \degree$, we can not compare directly their radius (and so their surface) to those of known sources (simulated or catalogued). So for each known source we introduce an effective radius as $R_{\it eff}=\sqrt{R_{source}^2+R_{corr}^2}$. We define $R_{source}$ as the outer radius for shell-like sources, the $1\sigma$ width for Gaussian-like sources, the radius for disk-like sources, and zero for point-like sources.

In the following we set $\rm SF_{ overlap}>0.3$ as association criterion. We also enforce that each source can be associated to only one object and conversely. This association criterion is used to estimate the fraction of simulated sources associated to a detected object (reconstruction fraction, see Sect.~\ref{sec:SimTest}), and the fraction of objects associated to a known source (association fraction, see Sect.~\ref{sec:discu}).

We have tested associations with the HGPS catalog sources and their Gaussian components separately. Correlations in position and radius of detected objects with HGPS sources are further discussed in Sect.~\ref{sec:HGPScomp}. Additionally, based on the information provided by the 4FGL, SNRcat and \textit{Fermi}-LAT SNR catalogues, we report on possible MW associations for each object.
%Independently of this association procedure we add a flag for the detected objects whose centers are within the extraction region for an HGPS source, even if they are not closely matching the position of the source or one of its Gaussian components (if the extraction region parameter is not available we use the source size). 

\begin{figure*}
\centering 
\includegraphics[scale=0.8]{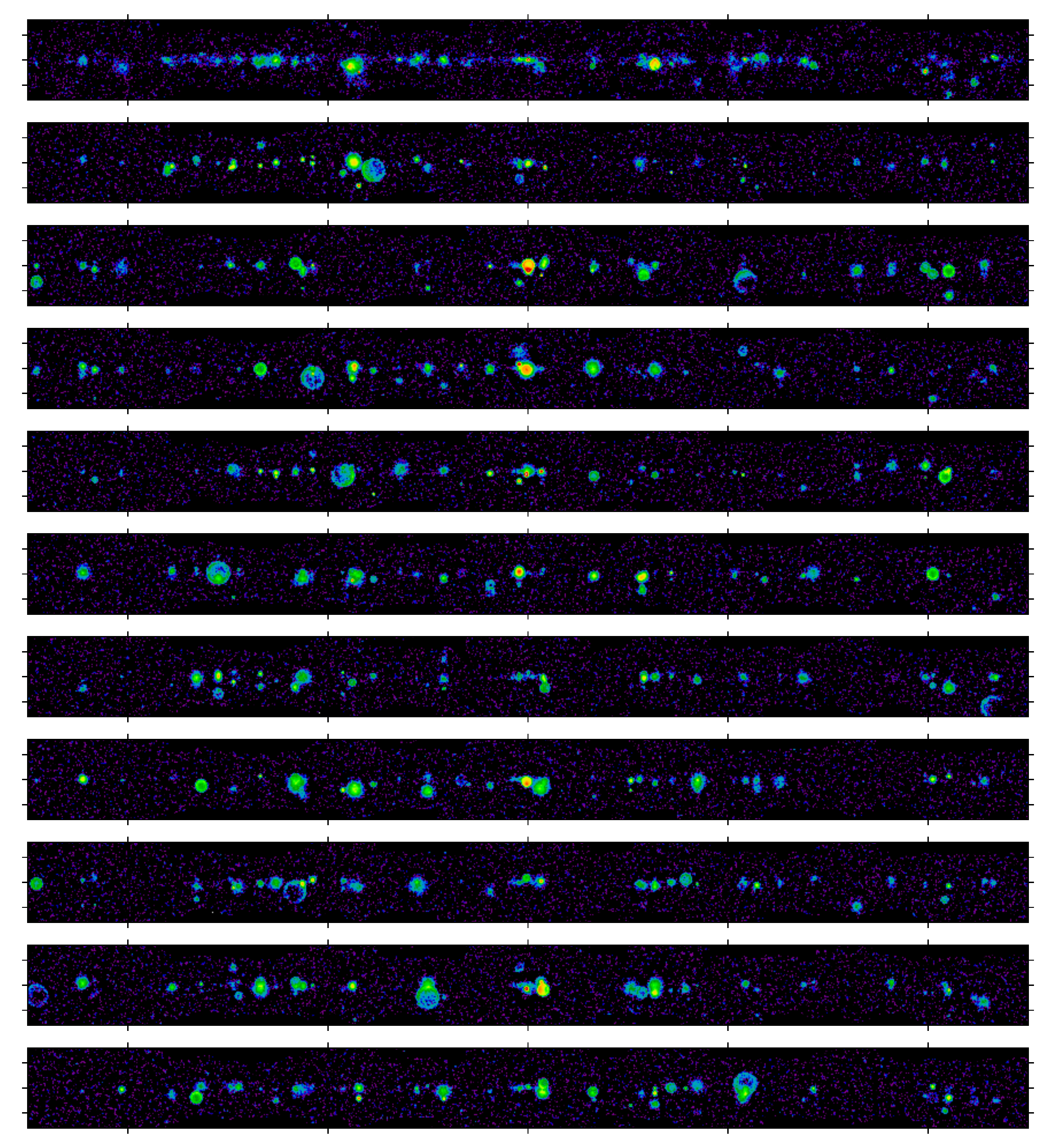}  
\caption{Significance maps sample -- Top row: Significance map from the HGPS survey. Other rows: Significance maps for 10 simulated skies, each one including 100 sources and a toy-model for background emission (with $C_{ism}=0.1$, see~\ref{apx:Sim_data}). In all panels, significance maps are derived using the Adaptive Ring Background method with R$_{corr}$ = 0.1\degree. We display only $|l|<50\degree$ but simulations have the same coverage as the HGPS survey.}
\label{fig:Signi_sample}
\end{figure*}

\begin{figure*}
\centering   
\includegraphics[width=\hsize]{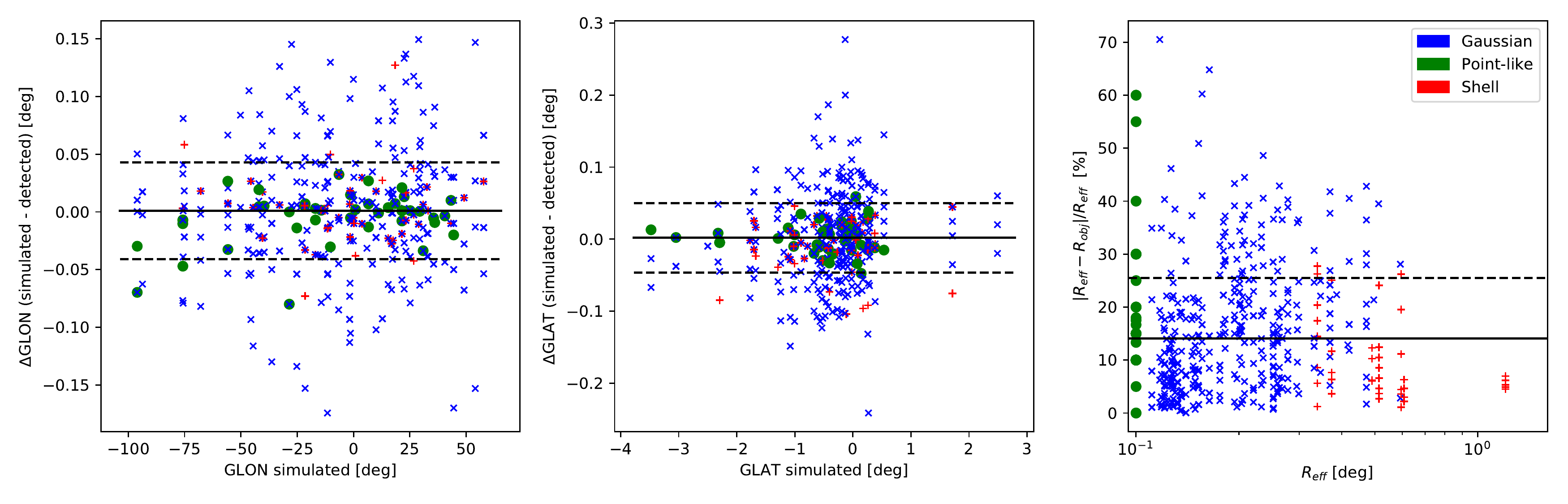}
\caption{Positional correlations between detected objects and  simulated sources -- The left and middle panels give the differences in longitude and latitude, respectively. The right panel shows the absolute relative error on radius. Note that the effective radii, $R_{\it eff}$ given in the right panel are the quadratic sum of the simulate radii and the 0.1\degree~correlation radius of the significance map used to determine the Hough circle radii, so their values can be readily compared. In each panel the mean and standard deviation values are given by the plain and dotted black lines, respectively. The colors correspond to the morphological models used for the simulated sources.}
\label{fig:corrSIM}
\vspace{2cm}
\begin{minipage}{0.47\hsize}
\centering 
\includegraphics[width=\hsize]{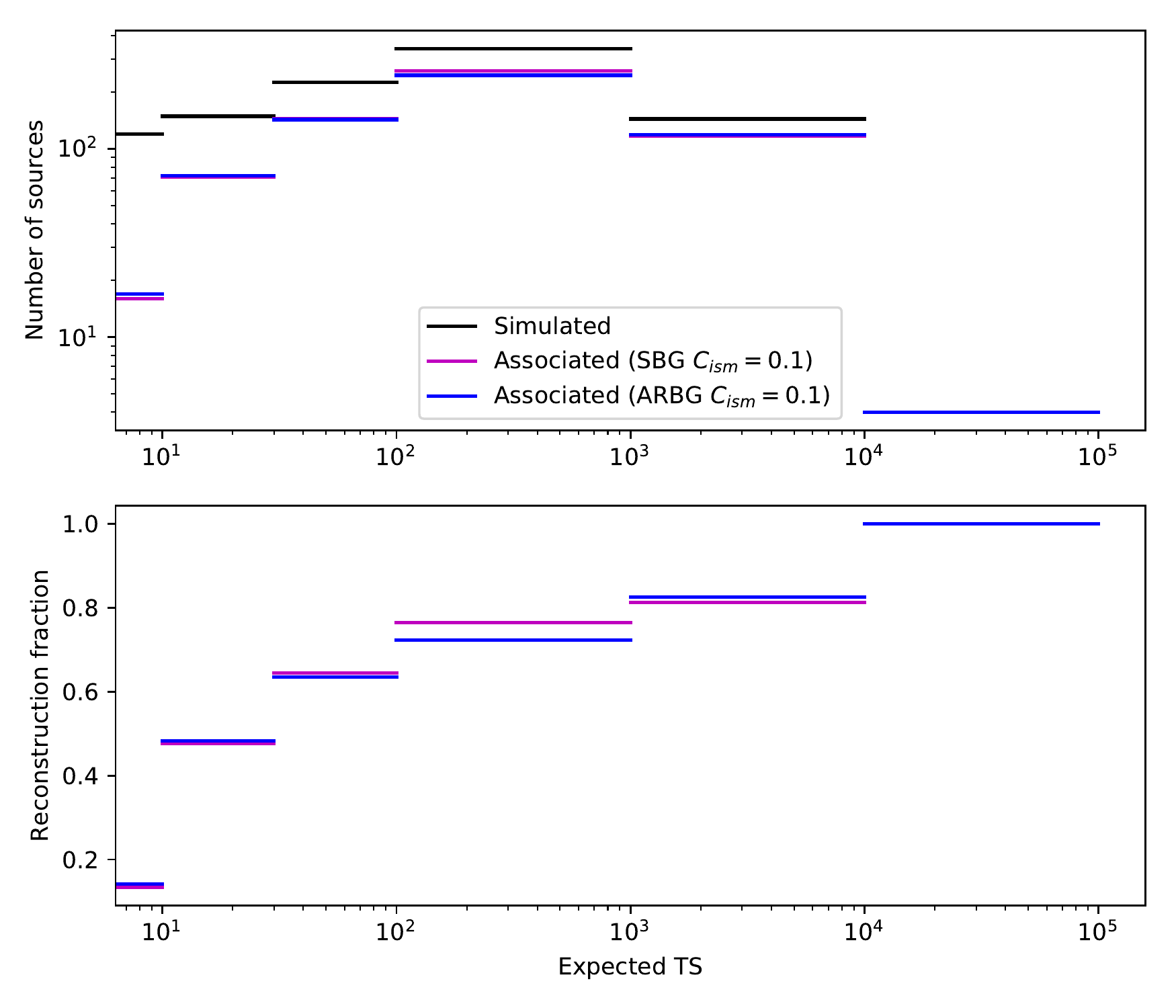}  
\caption{Reconstruction fraction as a function of the expected TS -- Top: Histogram of the number of simulated sources within bins in expected TS values. Bottom: Fraction of simulated sources associated to a detected object in the same bins in expected TS values. Magenta and blue bars correspond to objects detected as Hough circles in the significance map derived using simulated background (SBG) and through the adaptive ring background method (ARBG), respectively.}
\label{fig:Sim_detfracTS}
\vspace{1cm}
\end{minipage}
\hspace{0.7cm}
\begin{minipage}{0.47\hsize}
\centering 
\includegraphics[width=\hsize]{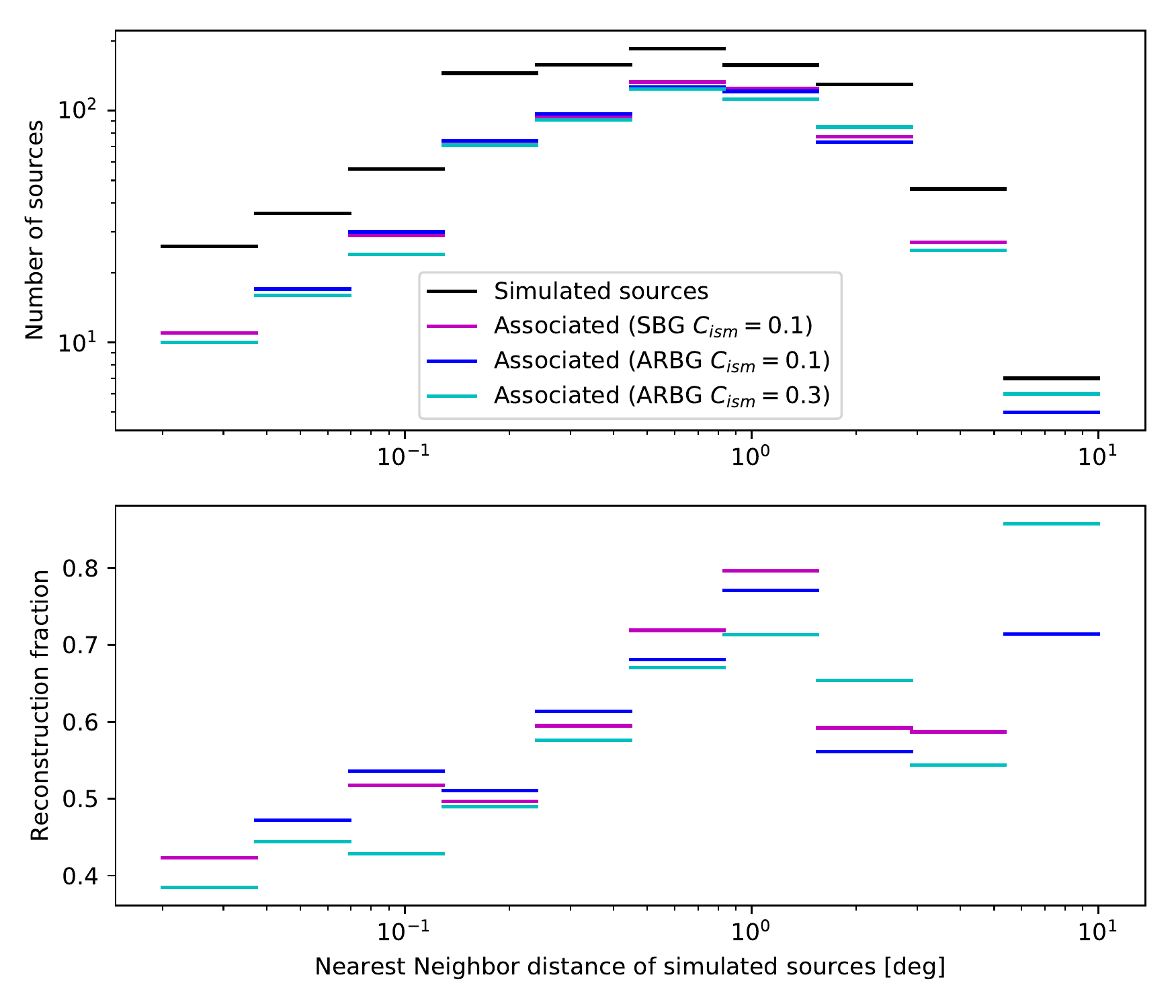}  
\caption{Reconstruction fraction as a function of the nearest neighbor distance d$_{nn}$ -- Top: Histogram of the number of simulated source within bins in d$_{nn}$. Bottom: Fraction of simulated source associated to a detected object in the same bins in d$_{nn}$. Magenta and blue bars correspond to objects detected as Hough circles in the significance map derived using simulated background (SBG) and through the adaptive ring background method (ARBG), respectively. Cyan bars correspond to the ARBG case with a higher diffuse background level}.
\label{fig:Sim_detfracDNN}
\vspace{1cm}
\end{minipage}
\end{figure*}

\section{Tests on simulations}
\label{sec:SimTest}

\subsection{Simulated datasets} 
\label{sec:Simdata}

We have performed a bootstrap of the sources catalogued in the HGPS survey. We produced 10 sky realisations, each one including 100 sources, a model for hadronic background emission  and a toy-model for diffuse Galactic background emission. Details on the production of the simulated flux, counts, exposure and final significance maps are given in \ref{apx:Sim_data}. We did not intend to produce a realistic model for the various components of the diffuse Galactic emission, instead we used a simplistic toy-model that we can tweak in order to test how the source reconstruction is affected by an increase of the diffuse background level. To do so, we arbitrarily tweaked the contrast in flux between the source and the diffuse background with one parameter. The contrast parameter $C_{ism}$ has been set to 0.1 in the minimal case and 0.3 for a stronger diffuse emission. 
In order to test the systematics related to background estimation, we produced significance maps using the exact simulated background model (noted SBG in figures) or through the adaptive ring background method as defined in the HGPS (noted as ARBG in figures). We show in Fig. \ref{fig:Signi_sample} the H.E.S.S. significance map and 10 simulated ARBG significance maps in the minimal diffuse emission case ($C_{ism}=0.1$). We illustrate the application of the method on one of the simulated sky in Fig. \ref{fig:mth_ex}, \ref{fig:Sim_preprocess} and \ref{fig:Sim_sources_comp}.

\subsection{Object detection performance} 

By design the detection algorithm will find more structures than there are sources. So its performance depends on the capability in reconstructing the sources correctly, among the sub-structures.
We define the reconstruction fraction as the fraction of simulated sources associated to detected objects using the procedure introduced in Sect.~\ref{sec:xmatch}.
The association criterion based on inter-center distance and surface overlap are strong enough to ensure a close match in position and extension between the simulated sources and detected objects (as shown in Fig. \ref{fig:corrSIM}). The differences in the longitude, latitude and radius have a mean value compatible with zero with a root-mean-square dispersion (rms) lower than $0.05\degree$. The difference in radius increases with radius, but the absolute relative error is mostly stable around its average value of $\sim$15\%.

The detectability of \g-ray sources is usually expressed in terms of their statistical significance.
For each simulated source we can estimate their expected test statistic (TS) as the log-likelihood ratio between the exact simulated model and the same model excluding the source. In Fig. \ref{fig:Sim_detfracTS} we show that the reconstruction fraction of simulated sources increases with their expected TS, and exceeds  70\% above TS$>30$.  
Fig. \ref{fig:Sim_detfracTS} also shows that the use of the adaptive ring background rather than the perfect simulated background in the significance map calculation changes the reconstruction fraction by only a few percent, even when we considered a higher diffuse background level (Fig. \ref{fig:Sim_detfracTS3}).

In order to quantify the effect of the source confusion on the source reconstruction capabilities, we plot in Fig. \ref{fig:Sim_detfracDNN} the reconstruction fraction as a function of the nearest neighbour distance of the simulated source, d$_{nn}$. The reconstruction fraction progressively increases from 40\%~to 80\% for sources closer than $1\degree$. For the closest sources the low fraction is primarily explained by the source confusion while the effect of the diffuse background level is secondary.  
For sources with d$_{nn}$ larger than $1\degree$~and with an expected TS$>30$ the reconstruction fraction is stable around 80\% while it drops if we also consider sources with lower TS (see Fig. \ref{fig:Sim_detfracDNNTS}). % The source distant by few degree lies mostly at higher latitude or in the outer Galaxy where the exposure of the survey is lower so their signal to noise ratio is lower. 

Note that we generated the simulated source sample from the detected source distribution and we did not extrapolate the distribution to include sources with lower flux, so the effect of source confusion is potentially underestimated. The reconstruction fraction will likely be lower for the low TS values (see the two first bins of Fig  \ref{fig:Sim_detfracDNN}).% The sources with higher TS values should be less affected, as in that case the source confusion effect would likely be driven by close enough sources with equivalent or higher flux.  

\begin{figure*}[!p]
\centering 
\includegraphics[scale=0.515]{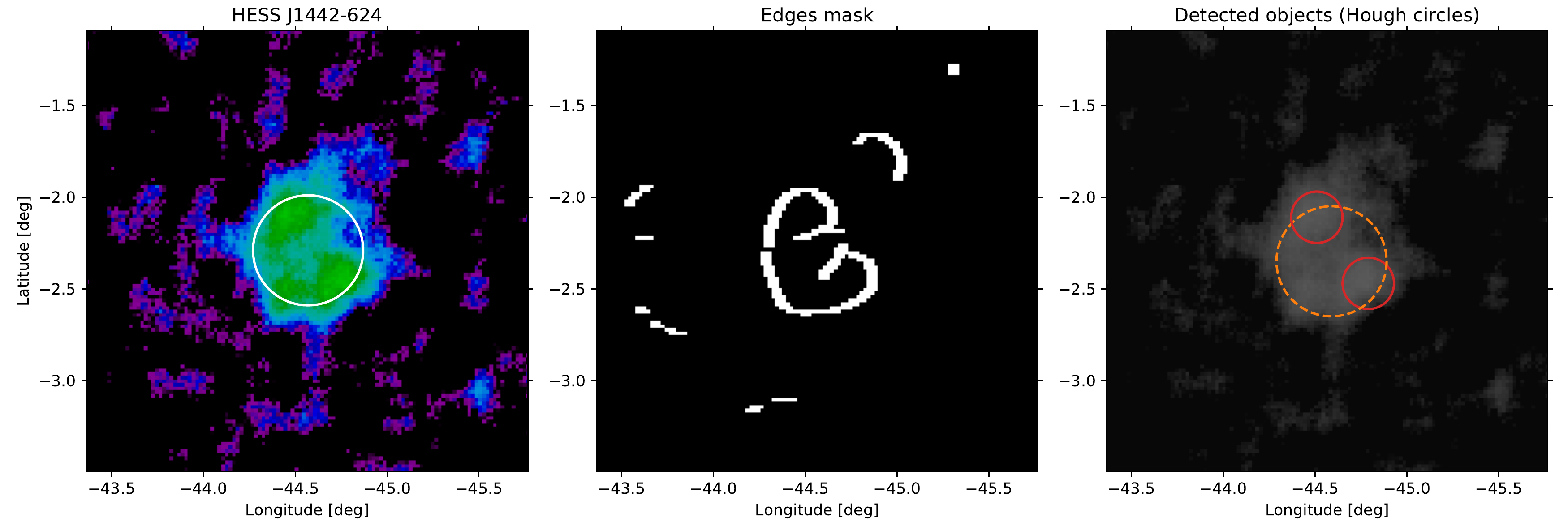}  
\includegraphics[scale=0.515]{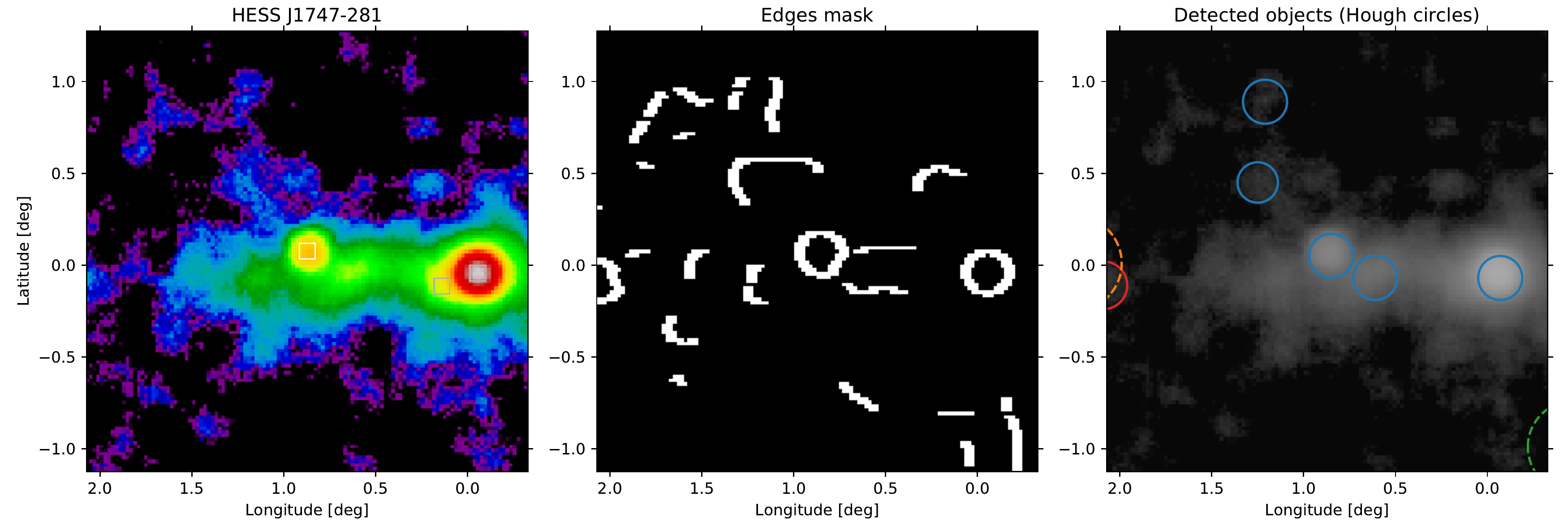}
\includegraphics[scale=0.515]{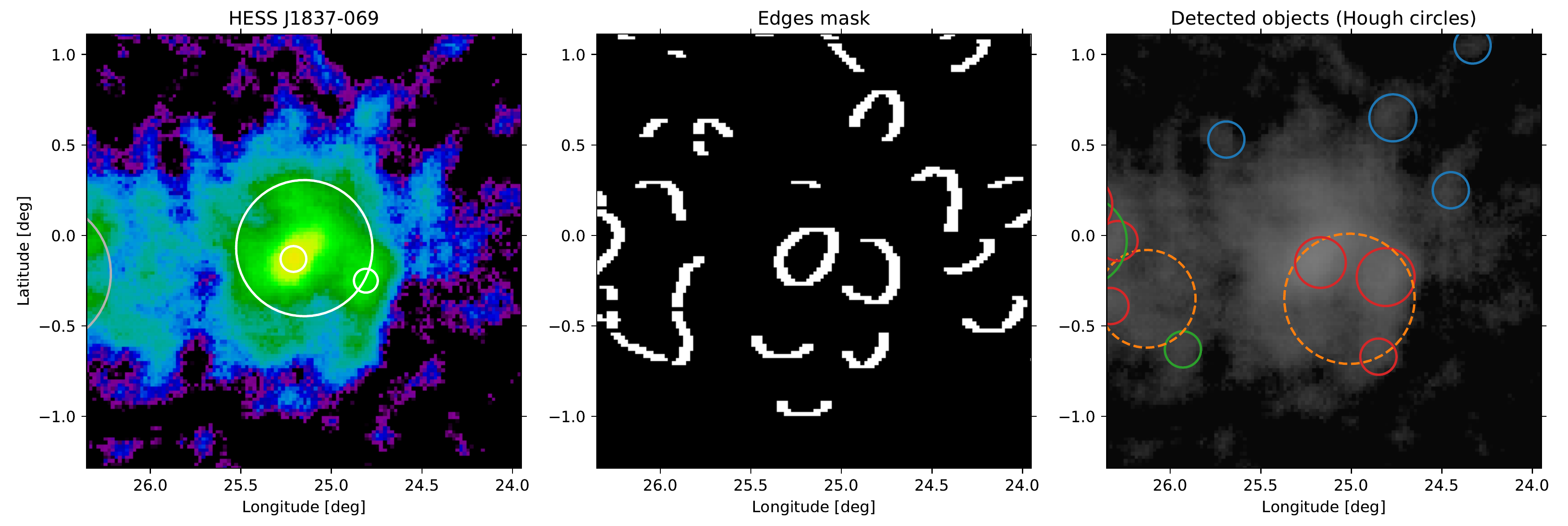}
\caption{H.E.S.S. sources as seen through edge detection and Hough circles -- Left column: Significance maps (R$_{corr}$ = 0.1\degree) and catalogued source components from the HGPS (circles correspond to the sizes of the Gaussian components if available or to the source radii otherwise, squares indicate point-like sources; the source given in the panel title appears in white and the others in grey). Middle column: Edge masks obtained after applying the Canny filter. Right column: Significance maps (same as the left column but in grey-scale) and objects detected; dashed circles denote cavities or flat regions, plain circles represent peaks; blue circles flag isolated objects (class 0), orange circles main-objects associated with sub-structures (class 1), and other colors correspond to the sub-structures (green, red, purple for class 2, 3, 4 respectively). Further information on classification flags is given in Sect.~\ref{sec:params}.}
\label{fig:Comparisons_HESS}
\end{figure*}

\begin{figure*}[!p]
\centering   
\includegraphics[scale=0.518]{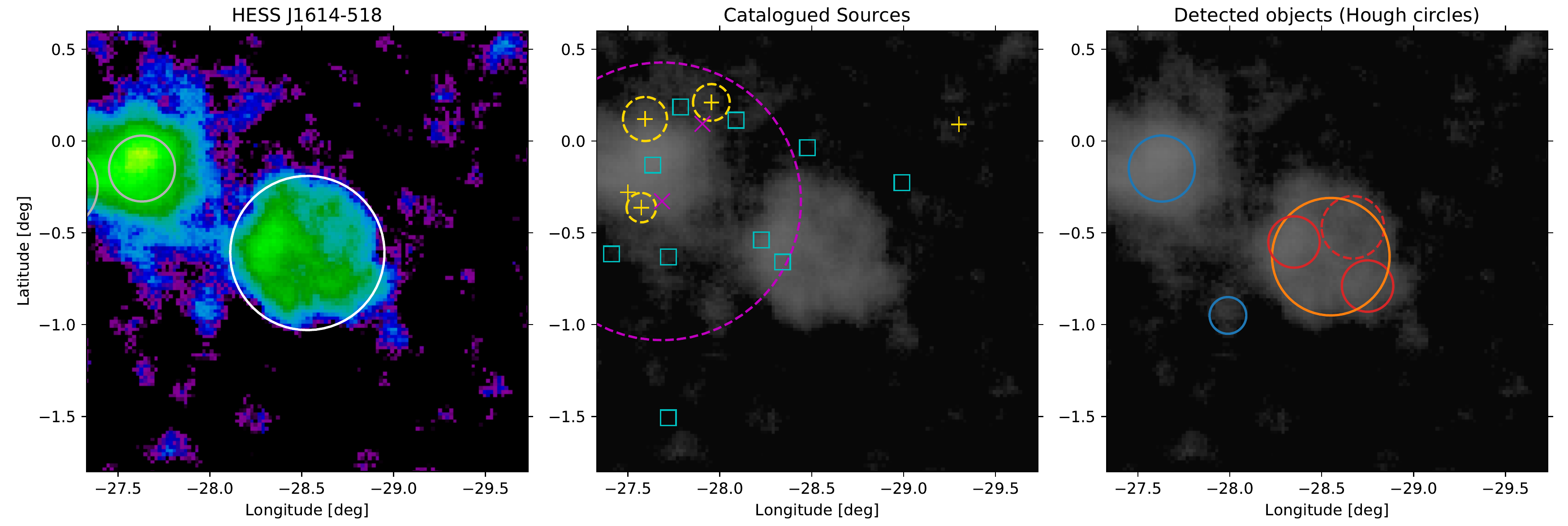}
\includegraphics[scale=0.518]{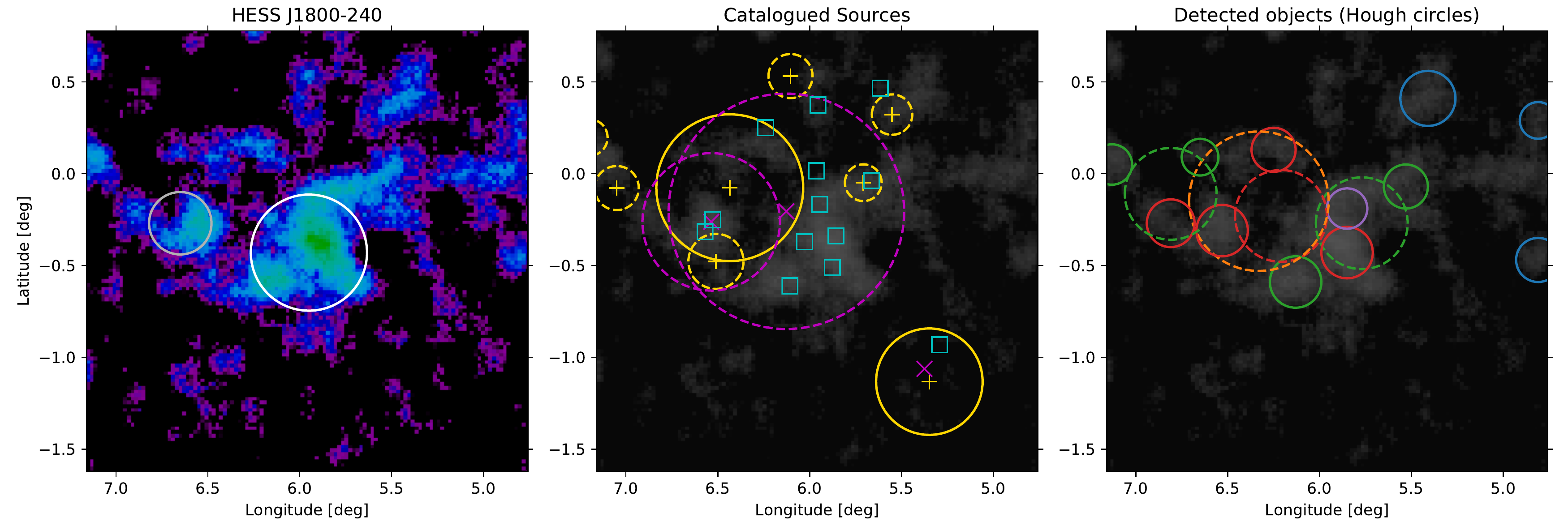}
\includegraphics[scale=0.518]{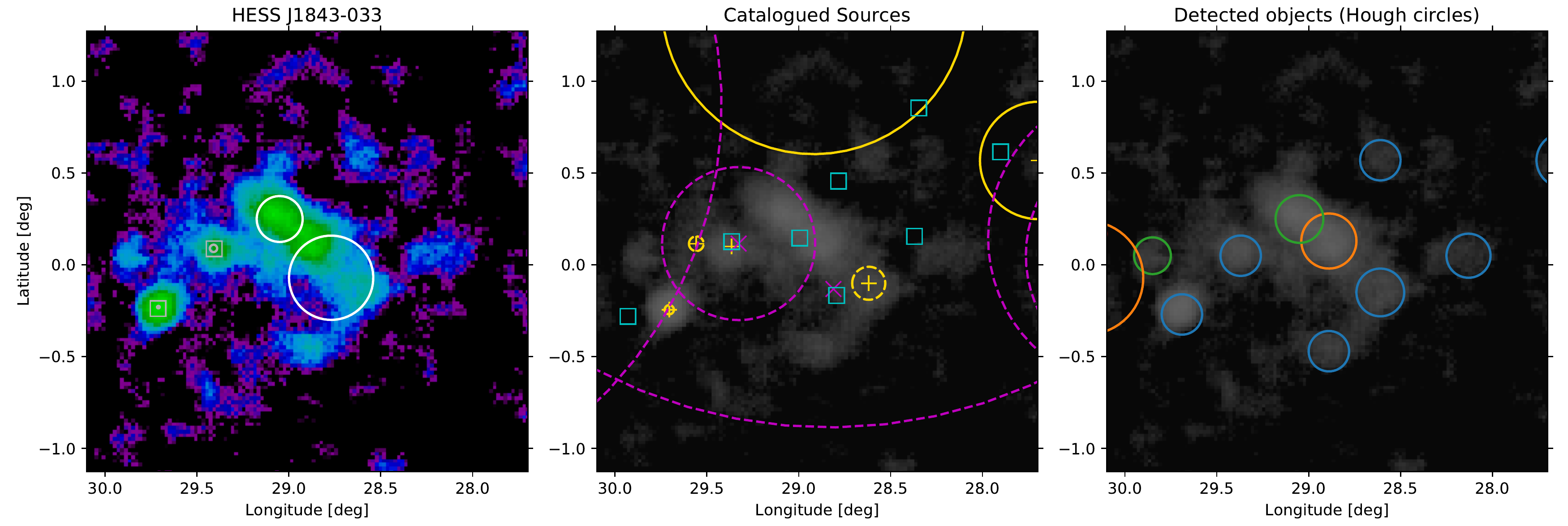}
\caption{Comparison between catalogued sources and detected objects -- Left column: Significance maps (R$_{corr}$ = 0.1\degree) and catalogued source components from the HGPS (same description as first column of Fig.~\ref{fig:Comparisons_HESS}). Middle column: Significance maps (R$_{corr}$ = 0.1\degree) and MW catalogued sources; blue squares are \textit{Fermi}-LAT sources from the 4FGL list; yellow items correspond to radio sources given by the SNRcat, circles give the radii when available, dotted circles correspond to SNRs \citep{2012AdSpR..49.1313F}; magenta items correspond to extended \textit{Fermi}-LAT sources from the SC1  \citep{2016ApJS..224....8A}. Right column: Significance maps and objects detected (same description as the last column of Fig.~\ref{fig:Comparisons_HESS}).}
\label{fig:Comparisons_CAT}
\end{figure*}

\section{Application to the HGPS survey}
\label{sec:HGPScomp}

%\subsubsection{Objects detected compared to the HGPS sources}

The object detection procedure was applied on the publicly available HGPS significance map with a correlation radius of 0.1\degree. The detection procedure results in a list of 462 objects among which 207 have centers inside the extraction radius of a HGPS source. Their spatial distribution across the whole region studied is shown in Fig.~\ref{fig:allsky}. We illustrate the detected objects compared to catalogued sources toward few smaller regions in Fig \ref{fig:Comparisons_HESS} and \ref{fig:Comparisons_CAT}.  Following the cross-match procedure described in Sect.~\ref{sec:xmatch}, 72 objects are coincident with a HGPS source within the inter-center distance limit (\mbox{$d_{c}<0.1+0.3\times R_{object}$}) and 64 also match the association criterion based on surface overlap ($\rm SF_{ overlap}>0.3$). Similarly, we have 74 objects close to a HGPS Gaussian component and 67 fulfilling the surface overlap criterion.

For each object associated to a HGPS source we flag its type according to the HGPS identification (PWN, composite, SNR or binary). However, we have differentiated two sub-classes of SNRs based on the information provided by the gamma-cat. We reduce the so-called SNR class to only resolved shell-like SNRs (HESS J0852-463, HESS J1442-624, HESS J1534-571, HESS J1713-397 and HESS J1731-347), while the SNR-MC class includes unresolved HGPS-SNRs (HESS J1718-374, HESS J1911+090), and sources coincident with a radio SNR or associated to a Molecular Cloud (HESS J1800-240, HESS J1801-233). This differentiation is essential for accurate morphological classification. 
It should be also noted that the ``composite'' type in the HGPS refers to sources for which the angular resolution does not allow for a separation of the PWN from the SNR shell, so that formally one cannot discriminate between the PWN and the shell as the dominant source of the VHE emission \citep{2018A&A...612A...1H}. Nonetheless, the VHE PWN population study \citep{2018A&A...612A...2H} considers most of these VHE sources to be PWNe based on physics arguments.  The HGPS sources morphologically classified as composites would then mostly be younger and more compact PWNe than those firmly identified as PWNe.

\begin{figure}
\centering 
\includegraphics[width=\hsize]{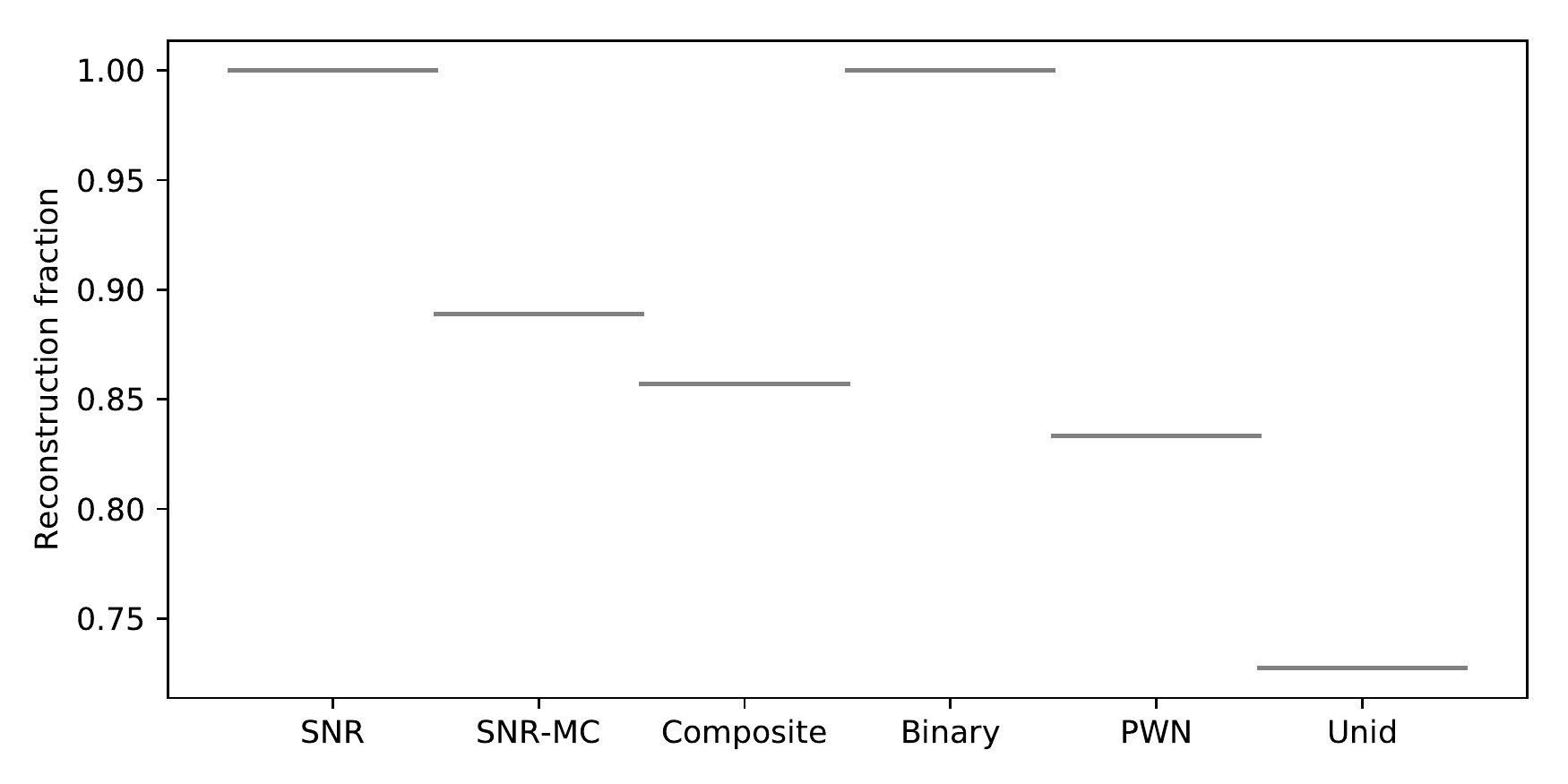}  
\caption{Reconstruction fraction with source types -- Fraction of HGPS sources associated to a detected object depending on their types (as given in the HGPS catalog).}
\label{fig:HGPS_detfrac_class}
\end{figure}

\begin{figure*}[p]
\centering   
\includegraphics[width=\hsize]{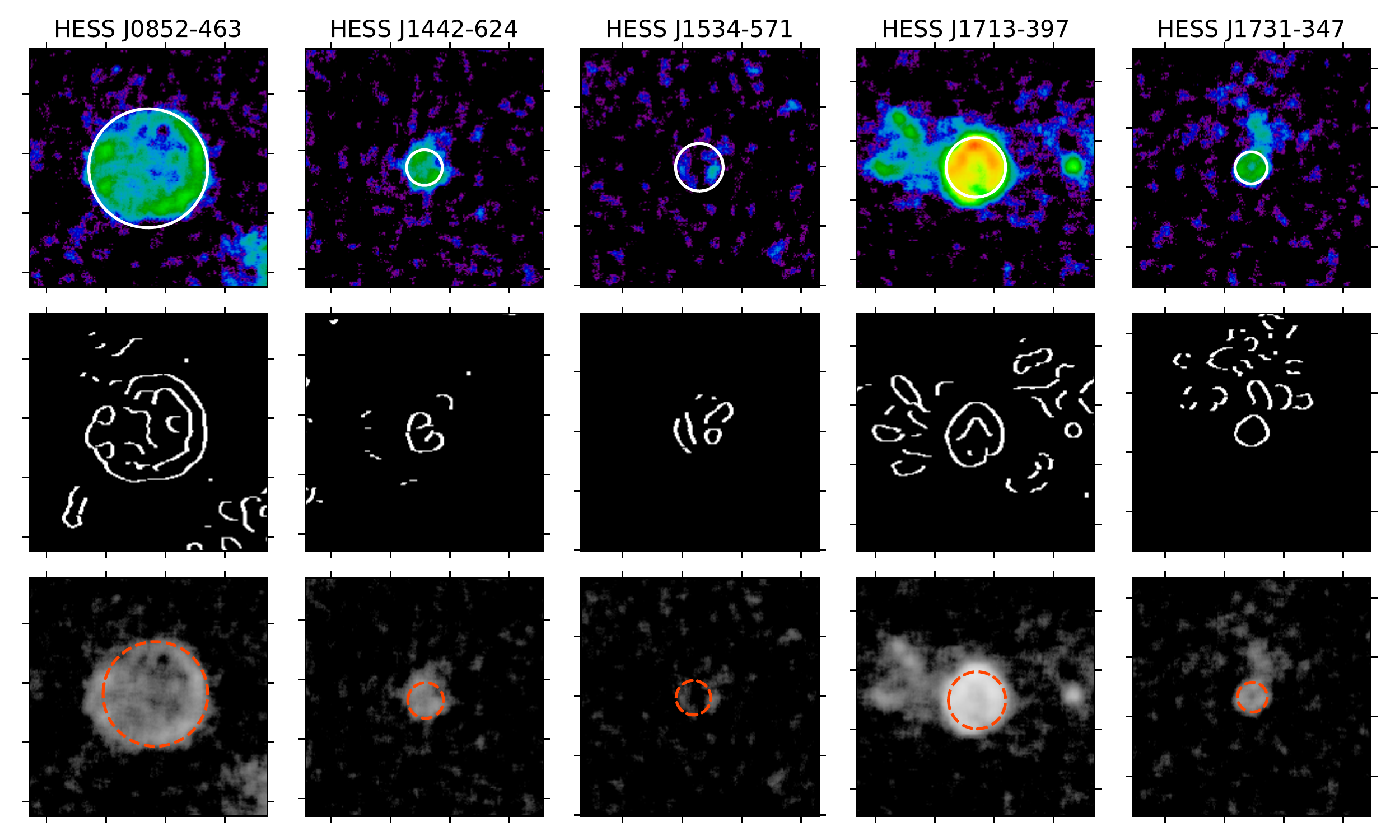}
\caption{Identification of the resolved shell-like SNRs -- First row: Significance map with R$_{corr}$ = 0.1\degree~and catalogued sources from the HGPS. Second row: edge mask obtained after applying the Canny filter. Third row: Significance map with R$_{corr}$ =0.1\degree~(same as first row but in grey-scale) and objects associated with the procedure described in Sec. \ref{sec:xmatch} (associations are unique so the eventual nested sub-structures related to those objects are not displayed here, but they are shown in Fig. \ref{fig:allsky}).}
\label{fig:SNRs}
\vspace{2.cm}
\centering   
\includegraphics[width=\hsize]{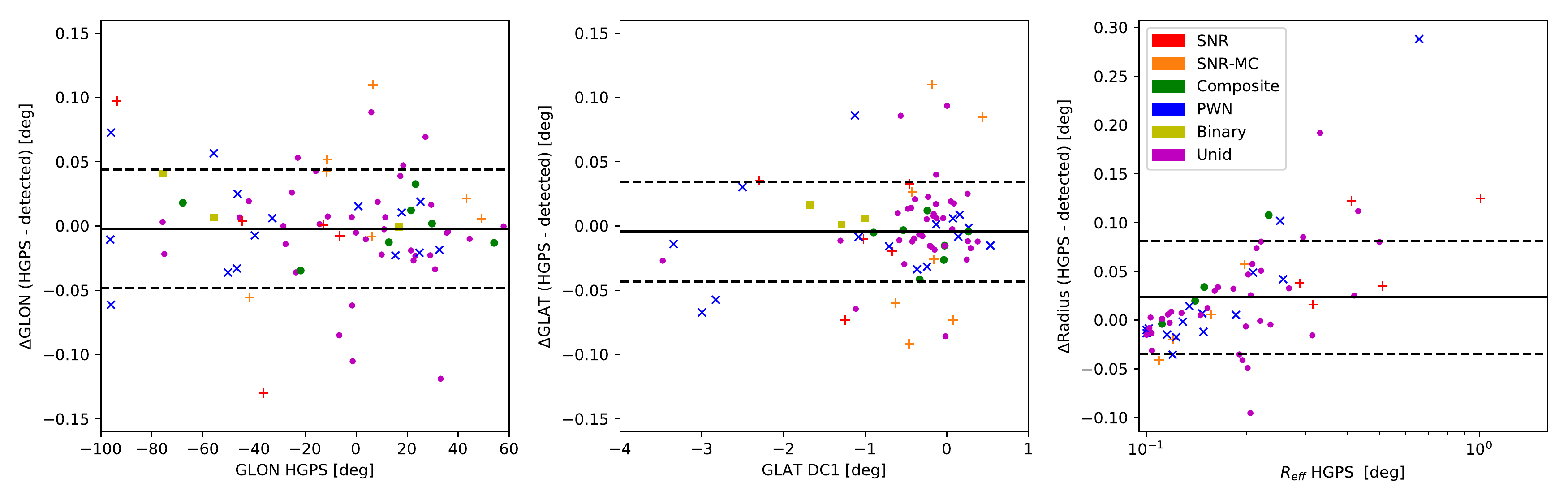}
\caption{Positional correlations between detected objects and  HGPS sources -- The left, middle, and right panels give the correlations in longitude, latitude and radius, respectively. Note that the effective radii,$R_{\it eff}$, given in the right panel are the quadratic sum of the catalogued radii and the 0.1\degree~correlation radius of the significance map used to determine the Hough circle radii, so their values can be readily compared. In each panel the mean and standard deviation values are given by the plain and dotted black lines, respectively. The colors correspond to the object types as given in the HGPS, except that we have differentiated the SNR-Molecular Cloud (MC) from the shell-type SNRs based on the information provided in the gamma-cat (see Sect.~\ref{sec:HGPScomp}).}
\label{fig:corrID}
\end{figure*}

In Fig. \ref{fig:HGPS_detfrac_class} we show the reconstruction fraction for the different source types catalogued in the HGPS. All the shell-type SNRs catalogued (based on dedicated analyses) are associated to at least one object. In Fig. \ref{fig:SNRs} we show the significance maps, their gradients and the edge masks used in the detection procedure toward the 5 well-resolved SNR shells in the HGPS. We found a good agreement in position and radius between the HGPS sources and the main objects detected, as shown in the first and last rows of Fig. \ref{fig:SNRs}.

The position and size correlations between the detected objects and the HGPS sources are shown in Fig. \ref{fig:corrID}. This figure demonstrates the excellent reconstruction in longitude and latitude. The error on the position given by the inter-center distance is 0.03\degree~on average with a standard deviation of $0.04\degree$. By construction we cannot find objects with radius lower than 0.1\degree~as we set the minimal radius in Hough space to the correlation radius of the significance map. We compare the detected radii with $R_{\it eff}$ given as the quadratic sum of the catalogued radii and the 0.1\degree~correlation radius. Most of the objects present a reasonably good correlation in radius within 20\% error, even though we note a systematic trend toward lower detected radii for sources larger than 0.2\degree~in radius.
These differences can be explained because the Hough transform tends to decompose a source in more sub-structures than the HGPS Gaussian components.

The HGPS components not associated to detected objects generally belong to one of the following cases:
\begin{itemize}
\item[-] Large Gaussian components may be decomposed in several sub-structures in the object list so there is no exact match (as seen in Fig. \ref{fig:Comparisons_CAT} for HESS J1843-033).
\item[-] Large Gaussian components associated to a smooth gradient are not seen because the edge detection imposes to preserve only sharp gradients (as seen in Fig.~\ref{fig:Comparisons_HESS} for HESS J1837-069). Thus, the object detection procedure is less sensitive to the diffuse background and to extended/faint PWNe.
\item[-] The 0.15\degree~limitation in the separation of objects with similar radius does not allow one to detect very close point-like sources; as an example, HESS J1745-290 in the Galactic center region is detected but not HESS J1746-285. In that case only the most prominent source is detected (because of a sharper significance gradient). This issue could be addressed by using another complementary technique, such as the object detection based on wavelet decomposition which is very efficient at isolating very close-by point-like objects from one another and from the underlying background \citep{1998AeAS..128..397S,2010A&A...517A..26S}.
\end{itemize}

%\begin{figure}
%\centering 
%\includegraphics[width=\hsize]{./img/SAHA_tst_sources_Reconstruction_fraction_TSfit.pdf}  
%\caption{Reconstruction fraction with fitted TS (HGPS Gaussians) -- Top: Histogram of the number of HGPS Gaussian sources within bins in fitted TS values. Bottom: Fraction of catalogued source associated to a detected object in the same bins in fitted TS values. We consider only HGPS Gaussians components as we don't have a fitted TS for the other sources.}
%\label{fig:HGPSG_detfracTS}
%\end{figure}

%\clearpage 

In Fig. \ref{fig:learning0} we represent the detected objects in three morphological parameter spaces. The distributions of the different source types tend to indicate that they are separable using our set of parameters. For example, sources identified as composites or binaries are close to the maximum circularity (unity) and the minimal radius ($0.1\degree$), while those identified as PWNe have mostly lower circularity and larger radius (see left panel). The well-resolved shell-type SNRs are clearly separated from all the other identified source populations by their Pearson-R (see middle panel). We also note that sources detected by the HGPS pipeline (which exclude well-resolved SNRs and several unidentified sources) populate a more restricted fraction of the parameter space than the detected objects. The HGPS detection procedure, based on an iterative detection of Gaussian components, does lead to a selection bias that can explain such a difference. However, due to the uncertain nature of detected objects, a fraction of the parameter space would correspond to spurious detection associated with sub-structures or background. In the following we discuss this issue and how we could identify the most valuable objects.

\section{Discussion on the objects identification}
\label{sec:discu}

\begin{figure}
\centering 
\includegraphics[width=\hsize]{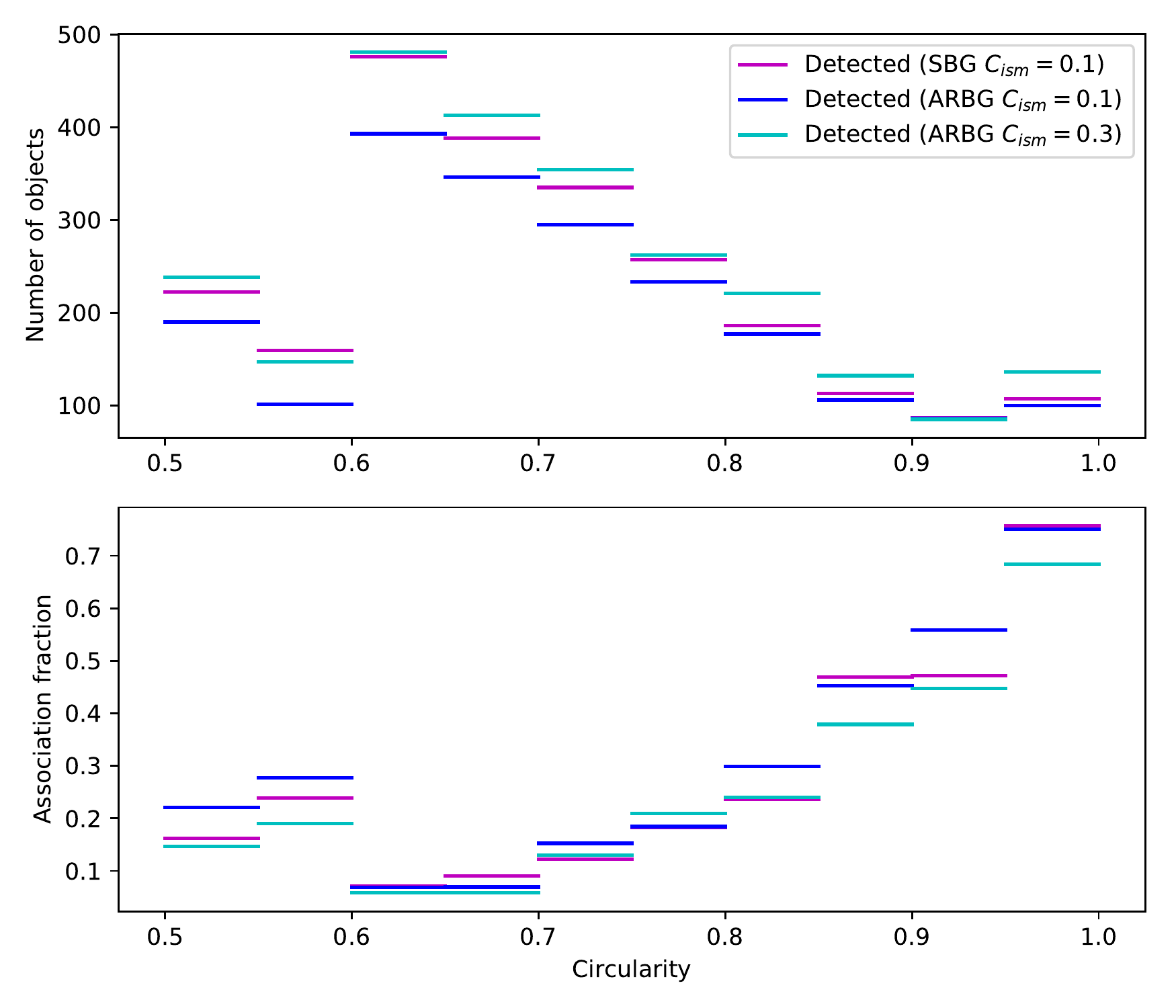}  
\caption{Association fraction (simulations) -- Top: Histogram of the number of sources within bins in circularity. Bottom: Fraction of detected objects associated to a simulated source in the same bins in circularity. Magenta and blue bars correspond to objects detected as Hough circles on the significance map derived using simulated background (SBG) and adaptive rings background (ARBG), respectively. Cyan bars correspond to the ARBG case with a higher diffuse background level.}
\label{fig:Sim_assofrac}
\centering 
\vspace{1cm}
\includegraphics[width=\hsize]{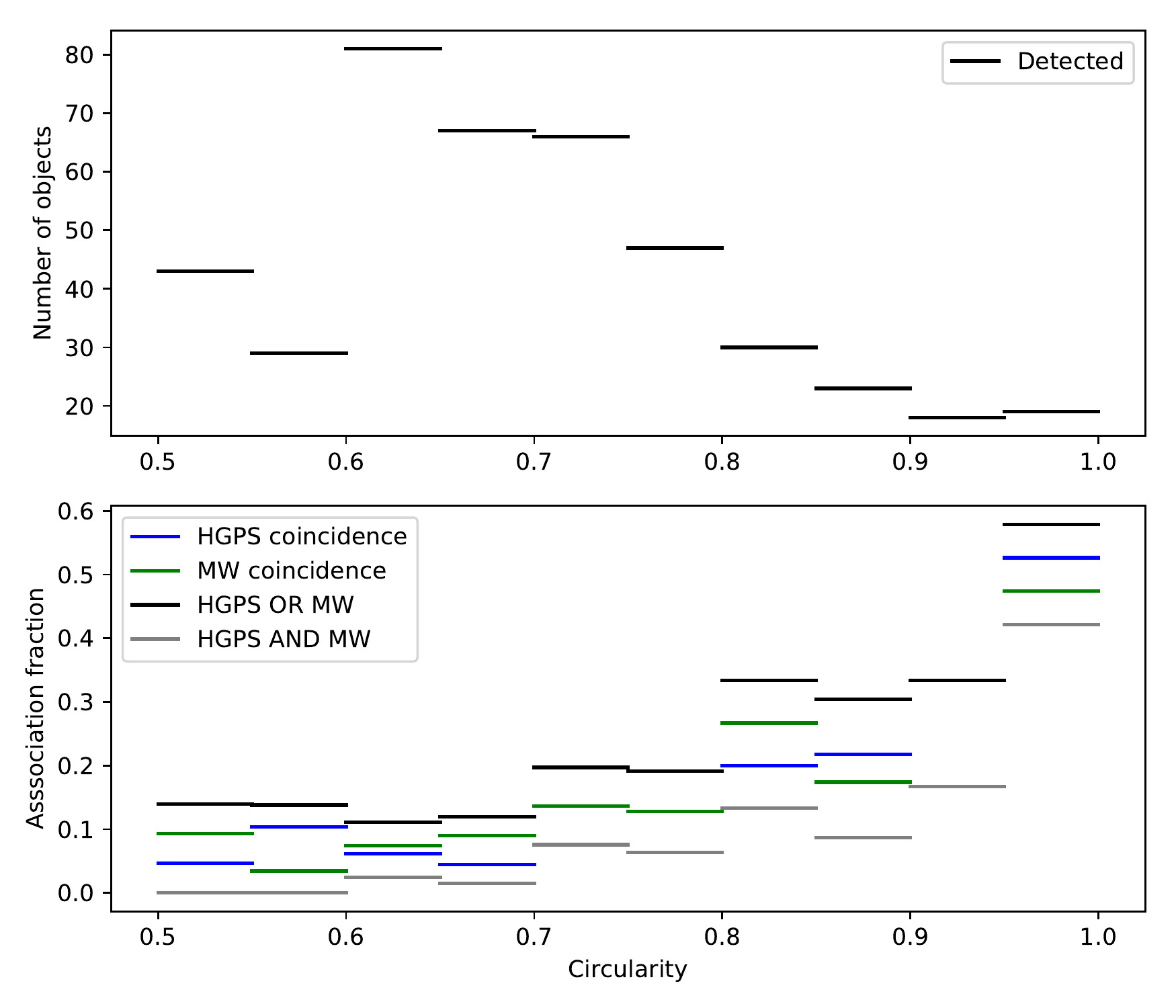}  
\caption{Association fraction (catalogues) -- Top: Histogram of the number of sources within bins in circularity. Bottom: Fraction of detected objects associated to a catalogued source in the same bins in circularity. Blue bars correspond to associations with the HGPS source or their Gaussian components. Green bars correspond to Multi-wavelength associations from the 4FGL, SNRcat or SC1 source catalogues.}
\label{fig:MW_assofrac}
\end{figure}

\begin{figure}
\centering 
\includegraphics[width=\hsize]{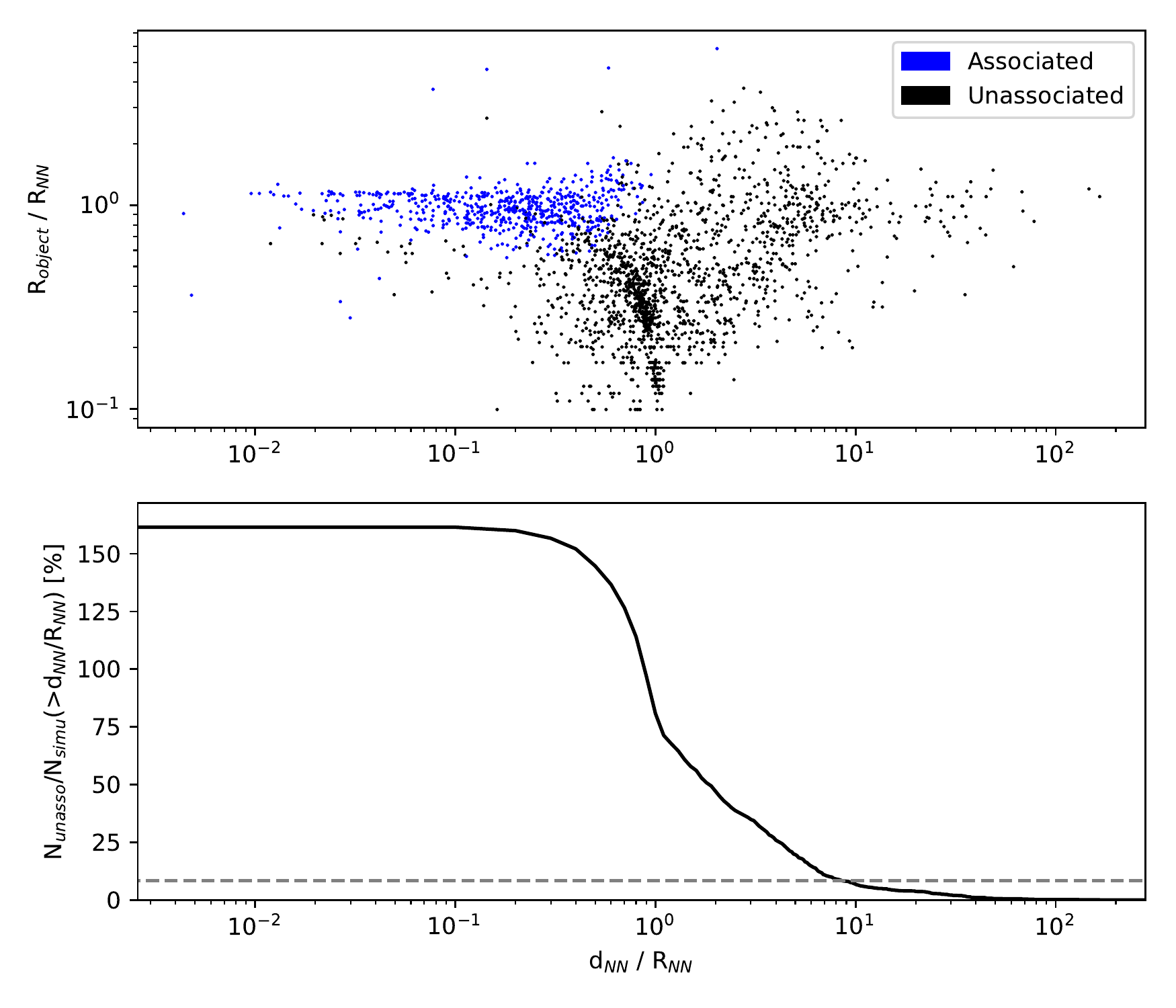}
\caption{Distance of the detected objects from simulated sources -- Top: Distribution of the associated (blue) and unassociated objects (black), the quantity on the horizontal 
axis is the distance of the object to the nearest simulated source in unit of source radius, d$_{NN}$/R$_{NN}$, and the quantity on the vertical axis is the ratio of the object radius to the source radius, R$_{object}$/R$_{NN}$. We show the results for the ARBG case (see Sect. \ref{sec:Simdata}).  Bottom: Anti-cumulative distribution of the unassociated objects with object-source distance (number of objects above a given value of d$_{NN}$/R$_{NN}$ per 100 source simulated). The dotted line is the average fake detections rate of noise-only simulations.}
\label{fig:dnn_rnn}
\end{figure}

\subsection{Disclaimer: not all detected objects are sources}
\label{sec:discl}

By design the object detection algorithm will find much more objects than there are sources, but this is expected for a preliminary list of seeds. For example, the latest \textit{Fermi}-LAT point source catalogues (4FGL or the preliminary FL8Y) contain about 5 000 sources while their initial detection step provides more than 13 000 objects \citep{2019arXiv190210045T}. According to our HGPS bootstrap simulations, we detect about 3.5 times more objects than simulated sources with an expected TS of 30 (or 3 times more if we consider sources with TS$>10$). In our case the main problem is that we have to deal mostly with nested extended objects. Among the nested objects, many will be internal sub-structures of the same source and not necessarily unrelated sources. Among the isolated objects, even if our detection procedure is by construction more sensitive to sources than background, several objects will be spurious detections triggered by statistical fluctuations or background structures from the Galactic large-scale emission. 

We performed tests on simulations containing only noise and given the hysteresis threshold imposed on the significance maps we found on average only 8 fake objects detected per simulated sky (compared to about 250 detected on each baseline simulations containing 100 sources), so pure noise is marginal in the fraction of spurious detections. Additionally we produced significance maps without and with diffuse Galactic background for two different flux levels. For 100 sources simulated per sky we found on average 253 objects for the case without Galactic background, 224 for the minimal background case, and 265 when increasing the Galactic background flux by a factor of 3. According to Fig.  \ref{fig:Sim_detfracDNN} and \ref{fig:Sim_assofrac}, the reconstruction and association fractions tend to decrease when the diffuse Galactic background level increases, but only by 5\% on average.

In Fig. \ref{fig:dnn_rnn} we show that the unassociated objects are mostly located at the edge of the simulated sources and have radii smaller than those of their parent sources. Moreover the number of unassociated objects decreases with the distance to simulated sources. The fraction of unassociated objects beyond $\sim$ 8-radius away from the sources matches the average false detection rate of the noise-only simulations. So we expect that the spurious detections are mostly bad reconstructions due to source confusion or irrelevant sub-structures within, or large and faint wings of, extended sources.

In order to minimize the spurious detections we can filter the objects as to maximize the association fraction to simulated sources or MW-catalogues without impairing the reconstruction fraction. To do so we can adjust the minimal thresholds on circularity or significance as defined in the Sect.~\ref{sec:locmax}. Alternatively we can focus on a subset of objects: isolate the main ones from their sub-structures; select objects that share similar morphological parameters to the firmly identified sources; or search only for objects with a MW-association in at least one wavelength. In the following, we aim to classify and filter the objects in order to narrow down the list to the most valuable ones for further dedicated MW analyses and deeper VHE observations.

\begin{figure*}
\centering 
\includegraphics[width=\hsize]{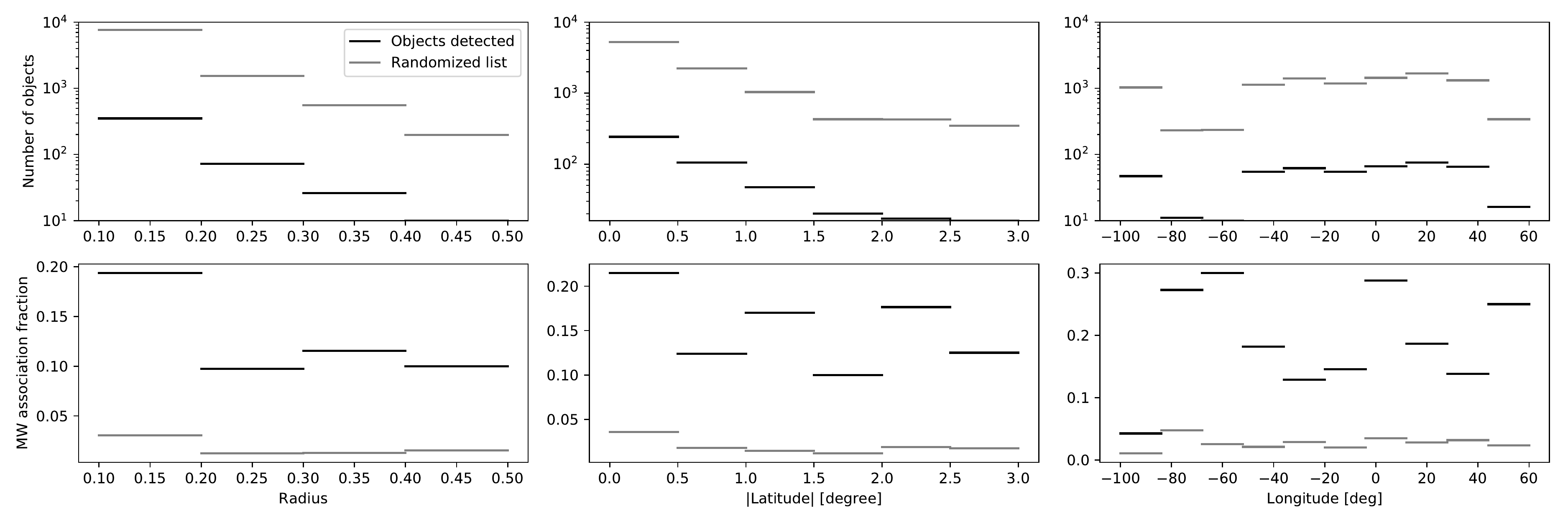}  
\caption{Multi-wavelength association fraction and coincidence by chance -- Top: Histogram of the number of objects within bins in longitude, latitude and radius (from left to right). Bottom: Fraction of objects associated to a 4FGL, SNRcat or SC1 source in the same bins. Black bars correspond to the list of objects detected in the HGPS significance map, and grey bars correspond to a randomized list generated by bootstrapping of the original list.}
\label{fig:Sim_assofrac_noise}
\centering 
\vspace{1cm}
\includegraphics[scale=0.7]{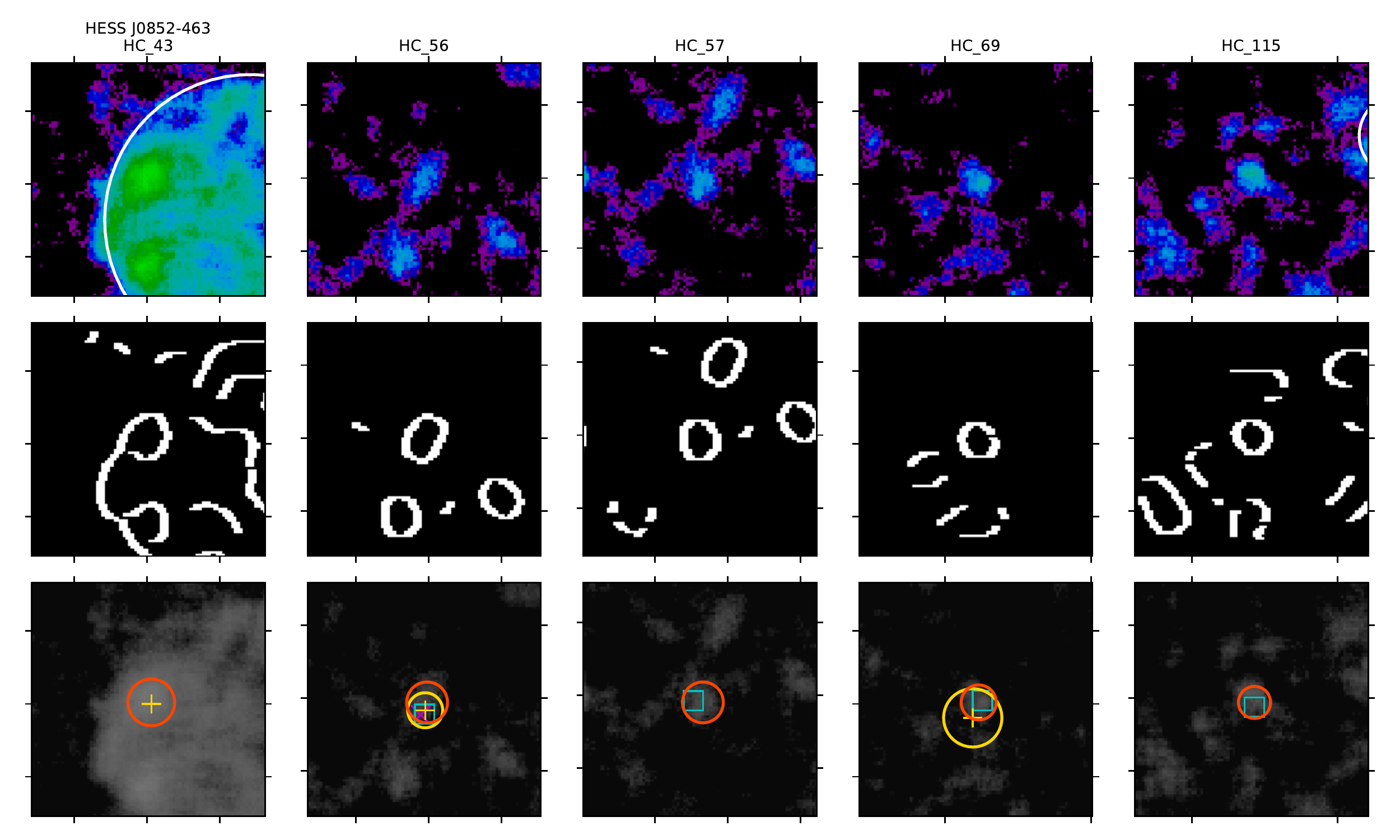}
\caption{Selection of high value targets based on circularity criterion and coincidence search -- First row: Significance maps (R$_{corr}$ = 0.1\degree) and catalogued sources from the HGPS. Second row: edge masks obtained after applying the Canny filter. Third row: Significance maps at R$_{corr}$ = 0.1\degree~(same as first row but in grey-scale). We select only objects with a circularity larger than 0.95 and coincident with a catalogued source in radio or GeV \g-ray wavelength (SF$_{overlap}>0.3$), but which are not catalogued in the HGPS. The detected objects are given as orange circles, blue squares are \textit{Fermi}-LAT sources from the 4FGL list; yellow items correspond to radio sources given by the SNRcat; magenta items correspond to extended \textit{Fermi}-LAT sources from the SC1.}
\label{fig:PScand}
\end{figure*}

\subsection{Candidate selection maximizing association fraction}
\label{sec:PScand}

We can flag valuable objects by searching for a MW-coincidence.
The selection can be further improved by identifying a subset of objects that maximize the ratio between the association fraction and the probability of coincidence by chance.

In order to quantify the fraction of coincidence by chance we have evaluated the association fraction to MW-catalogues of a randomized list of 10 000 objects. This list is obtained by bootstrapping the list of objects detected on the HGPS map, so the density of objects in longitudes, latitudes and radius are preserved (those quantities are independently drawn). In Fig. \ref{fig:Sim_assofrac_noise} we show the MW-association fraction to at least one of the sources from the 4FGL, SNRcat, or SC1 catalogues for the randomized list and the original one. Despite the large number of sources referenced in those catalogues the association fraction in the randomized list do not exceed few percent while the association fraction in the original one is about 10 times larger in average. This ratio can be further improved by selecting a subset of objects that show an enhanced association fraction. As shown in Fig. \ref{fig:Sim_assofrac} the fraction of detected objects associated to a simulated source increases with the circularity (\textit{i.e.} the fraction of the object circumference that triggered the detection). Note also that the condition we imposed to discard objects with circularity lower than 0.6 if their radius is lower than 0.25\degree~(see Sect.~\ref{sec:MthDet}) improves the association fraction at low circularity. The shape of this cut could be improved, but we prefer to avoid complex cuts by hand. The cut could be directly inferred from simulations in a more automated way using machine learning, and this also holds for the other parametric filters we chose for the source candidate identification. This is a wide topic to explore so we prefer to defer these refinements to future implementations. In Fig. \ref{fig:MW_assofrac} we show the fraction of detected objects associated with catalogued sources in the HGPS and/or at other wavelengths. We observe the same trend with an increasing association fraction with circularity.

Based on these observations we propose as valuable candidates a subset of objects with a circularity larger than 0.95 and associated to at least one source in the 4FGL, SNRcat or SC1 catalogues\footnote{These catalogues contain objects which belong to the two most representative TeV source classes, namely plerions (PWNe) and SNR shells. Nevertheless, these are certainly incomplete and other TeV emitters such as gamma-ray binaries or stellar clusters would thus be absent from our list of valuable candidates.} (but not catalogued in the HPGS). The 5 candidates resulting from this selection are shown in Fig. \ref{fig:PScand}. The content of this subset list is given in table \ref{tab:COMPcand}. We note that the \textit{Fermi}-LAT source coincident with HC\_57 candidate is catalogued as a pulsar (4FGL J1112.1-6108). The candidate HC\_56 is in coincidence with a plerionic composite SNR source (G291.0-0.1) and a \textit{Fermi}-LAT plusar (4FGL J1111.8-6039). The candidate HC\_69 is in coincidence with a plerionic composite SNR source (G308.8-0.1) and a \textit{Fermi}-LAT plusar (4FGL J1341.7-6216). Similarly the objects with circularity close to the unity that are associated to a HGPS source correspond mostly to small-size sources referenced as composites.

\begin{figure*}[p]
\centering   
\includegraphics[width=\hsize]{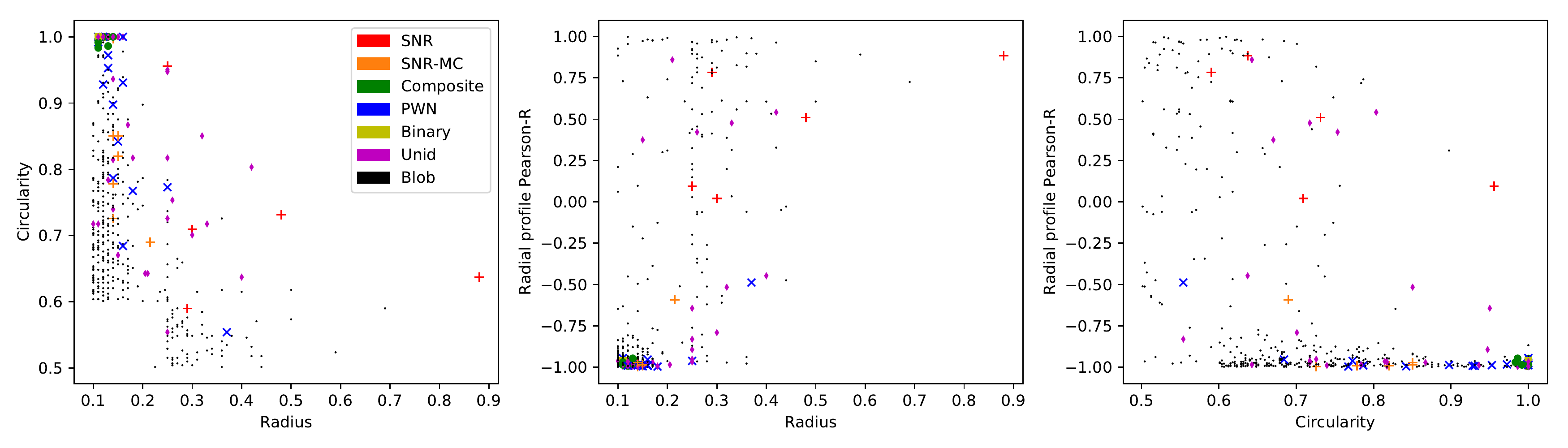}
\caption{Morphological parameters of the objects detected in the HGPS observations -- For each object we display the circularity \textit{versus} radius (left), the Pearson-R of the radial profile \textit{versus} radius (middle) and the Pearson-R versus the circularity (right). The colors correspond to the object type as given in the HGPS or gamma-cat (for the SNR-MC case). Black points correspond to detected objects with no HGPS counterpart (such that SF$_{overlap}<0.3$)}
\label{fig:learning0}
%, except that we have differentiated the SNR-Molecular Cloud (MC) associations from the shell-type SNRs based on the information provided in the gamma-cat
\vspace{0.2cm}
\centering   
\includegraphics[width=\hsize]{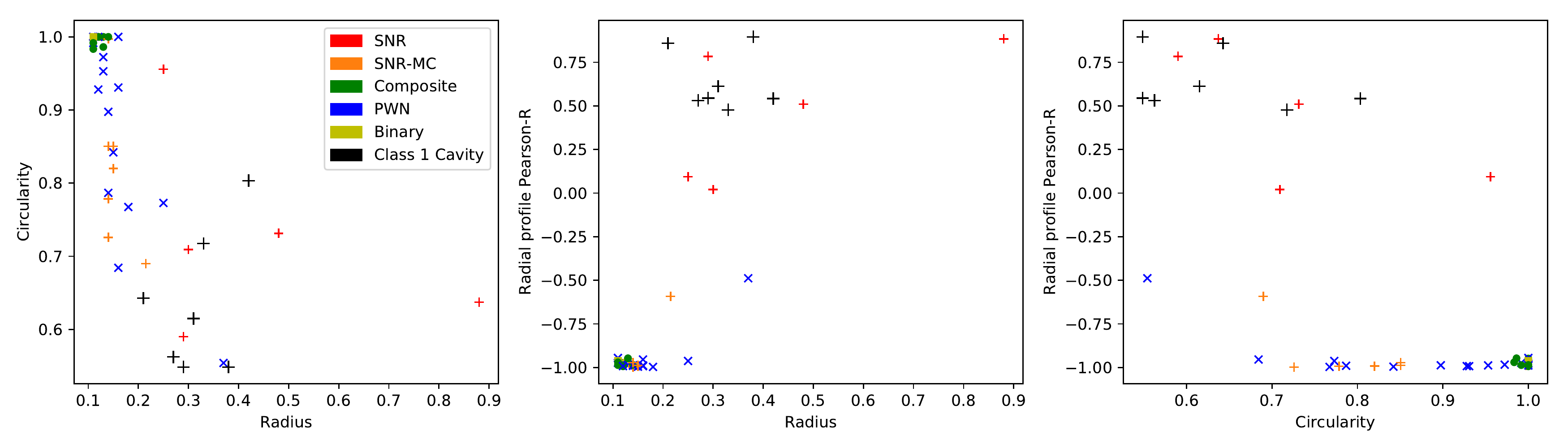}
\caption{Identification of shell-type SNR candidates: morphological classification  -- Same parameter space as in Figure \ref{fig:learning0}. We filtered the objects detected to select only class 1 cavities coincident with a catalogued source in radio or GeV \g-ray wavelength or an unidentified H.E.S.S. source (black plus). Class 1 cavities are defined as a main object with a Pearson-R larger than 1/3.}
\label{fig:SNRlearn}
\vspace{0.1cm}
\centering   
\includegraphics[width=\hsize]{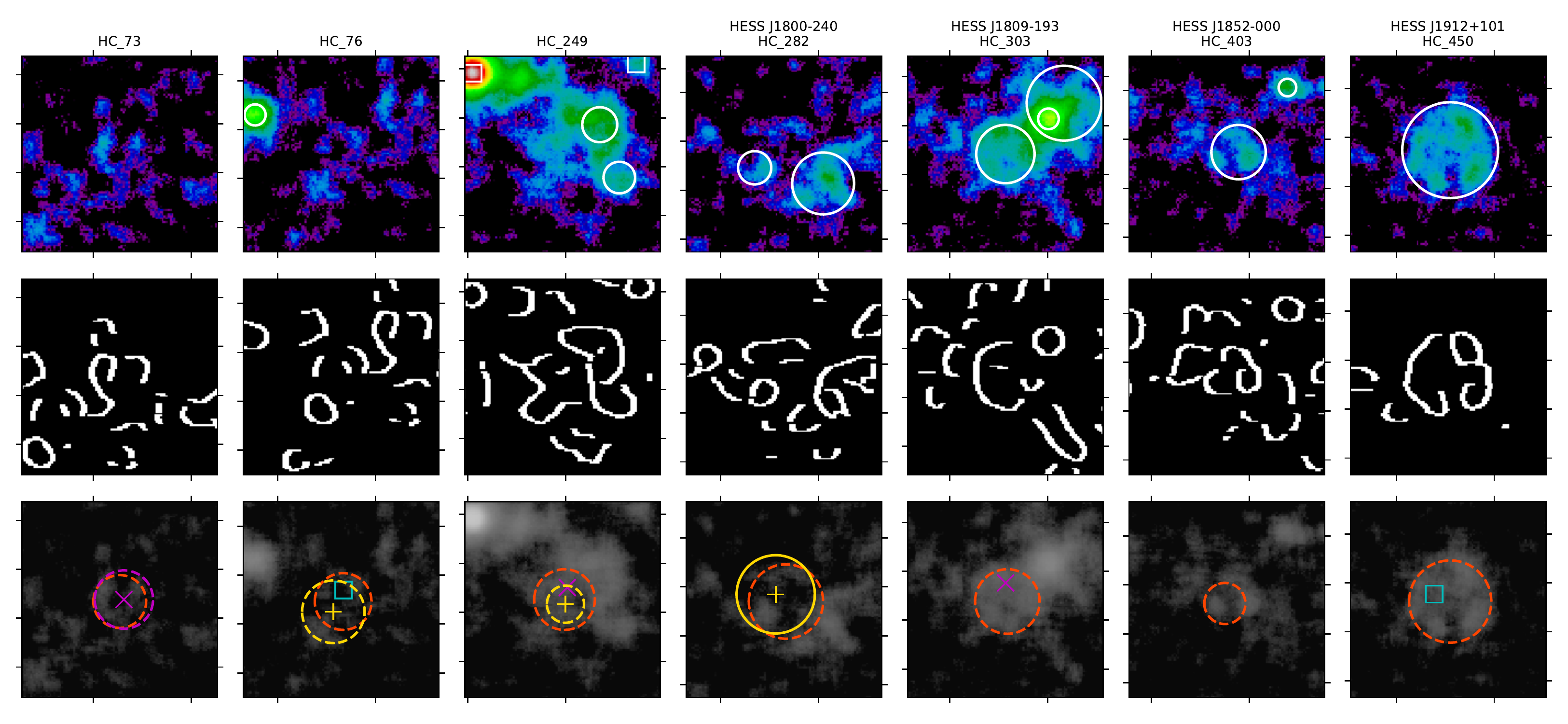}
\caption{Identification of SNR candidates: coincidence search -- First row: Significance maps (R$_{corr}$ = 0.1\degree). Second row: edge masks obtained after applying the Canny filter. Third row: Same as first row but in grey-scale;  We filtered the detected objects to select only class 1 cavities coincident with a catalogued source in radio or GeV \g-ray wavelength or an unidentified H.E.S.S. source with SF$_{overlap}>0.3$. The main-objects are displayed in orange (the nested sub-structures associated to the main-objects are not displayed here); blue squares are \textit{Fermi}-LAT sources from the 4FGL list; yellow items correspond to radio sources given by the SNRcat, circles give the radii when available, dotted circles correspond to SNRs; magenta items correspond to extended \textit{Fermi}-LAT sources from the SC1.}
\label{fig:SNRcand}
\end{figure*}

\subsection{SNR-like object identification}
\label{sec:SNRcand}

%\subsubsection{Morphology-based search}
%\subsubsection{Search for multi-wavelength counterparts}

In the simplistic object classification previously introduced, most of the well-known shell-like SNRs are identified as class 1 cavities (HESS J0852-463, HESS J1534-571, HESS J1713-397). In the parameter space that we defined, this class of objects corresponds to a main object with several nested sub-structures associated and with a radial profile exhibiting a Pearson-R coefficient larger than 1/3. Note that the two other known shell-like SNRs, namely HESS J1442-624 and HESS J1731-347, are classified as flat rather than cavities because the Pearson-R coefficient of the radial profile is close to zero as shown in Fig. \ref{fig:learning0}. Given their smaller size, the contrast between the outer ring and the inner part of the shell is less obvious (see Fig. \ref{fig:SNRs}). Filtering the list of detected objects to keep only class 1 cavities associated with a catalogued source results in 7 SNR-like candidates as shown in Fig. \ref{fig:SNRlearn}. In Fig. \ref{fig:SNRcand} we display maps of the significance, and the edge mask centered on these objects. Details on their properties are given in Table \ref{tab:SNRcand}.

Among these 7 SNR-like objects, 4 of them are already catalogued as unidentified sources in the HGPS \citep[HGPSC\_059 component of HESS J1809-193, J1800-240, J1852-000 and J1912+101,][]{2018A&A...612A...1H}, while 3 are found towards a (discarded) large HGPS structure (HGPSC 015 and HGPSC 051). In the HGPSC 015 region, we distinguish two objects : one (HC\_76) coincident with a radio SNR (G312.4-0.4) and one (HC\_73) with a \textit{Fermi}-LAT extended source, both showing a reasonably good match in radius with their potential counterparts (as illustrated in the two first rows of Fig. \ref{fig:SNRcand}). In the HGPSC 051 region, we report one object coincident with a radio SNR (G359.0-0.9) in the close vicinity of HESS J1745-303. Among the objects associated to unidentified HGPS sources, HESS J1912+101 has already been identified as a SNR candidate due to its shell-like morphology \citep{2018A&A...612A...8H}, and J1800-240 is potentially associated to the SNR W28 \citep{2008A&A...481..401A}. Interestingly in the W28 region the shell-like object is associated to the SNR while its substructures are associated to the surrounding HGPS sources. 

Four candidates show no evidence for association with a SNRcat source (including HESS J1912+101, see third row of Fig. \ref{fig:SNRcand}). Two candidates show partial matches, that is the main object is not coincident with a known shell but at least one of its nested sub-structures is coincident with a catalogued source in both radio and GeV \g-ray domains. Note that our MW-coincidence search is only based on catalogued sources, more information could be collected by looking directly at the MW images in order to propose additional observations in case of the presence of a potential, uncatalogued, counterpart.

We stress that focusing on Class 1 Cavities objects limits the results to SNR candidates for which the sub-structures are resolved. Less restrictive constraints could result in more candidates. For example, one of the sub-structures of the unidentified source HESS J1843-033, flagged as class 0, is coincident with G28.6-0.1 and can be considered as a poorly resolved SNR candidate (see the last row of Fig. \ref{fig:Comparisons_CAT}).

\section{Conclusion and perspectives}
\label{sec:ccl}

\subsection{Analysis products}
\label{sec:lists}

We provide the full list of objects detected on the HGPS significance map together with several short lists for convenience:
\begin{itemize}
\item[-] list of the objects associated with the HGPS sources (providing a detailed description of their sub-structures)
\item[-] list of non-HGPS objects with MW coincidences (which could be further studied in dedicated analyses)
\item[-] list of high-circularity candidates (see Sect.~\ref{sec:PScand})
\item[-] list of SNR-like candidates (see Sect.~\ref{sec:SNRcand})

\end{itemize}

In addition to the object positions (in longitude and latitude), these lists contain the morphological information defined in Sect.~\ref{sec:params} and the possible association given by the catalogue cross-match described in Sect.~\ref{sec:xmatch}. All these lists are available in \textit{FITS} format in the paper data on the editor page\footnote{\url{https://www.sciencedirect.com/science/article/abs/pii/S0927650520300347}}. A description of the \textit{FITS} file columns is provided in Table \ref{tab:content} in Appendix. Some examples of \textit{FITS} file contents are provided in Tables \ref{tab:COMPcand} and \ref{tab:SNRcand} (see Sects.~\ref{sec:PScand} and \ref{sec:SNRcand}).

\subsection{Summary and future works}

We have shown that using pattern recognition techniques we can extract pertinent structural information from the HGPS significance maps and detect objects with partial spherical symmetry without assuming a morphological template (e.g.\ point-like, Gaussian-like, or shell-like). In the approach proposed, sparse structural information is extracted using a edge detection operator. We then apply a Hough circle transform and detect a collection of objects as local maxima in Hough space. On the basis of morphological parameters we can characterize different object classes. Morphological classification and comparison with known source populations allow one to constrain the nature of the detected objects. In particular we have shown that the well-resolved shell-type SNRs catalogued in the HGPS can be isolated from the other source types with simple classification rules (Main-object with sub-structures and Pearson-R larger than 1/3). By extension, we can identify new source candidates that share the characteristics of the well-known sources. Moreover, catalogue cross-matches indicate that several detected objects not catalogued in the HGPS are spatially coincident with MW counterparts. Those are more likely to not be coincidence by chance when considering objects with higher circularity.

We have presented a prospective study on the use of pattern recognition and classification techniques to detect source-like objects in the VHE \g-ray sky and identify new valuable source candidates. We have no doubt about the potential of this approach, but there is still a lot to explore as shown by the various perspectives discussed throughout this paper.  We list other possible technical improvements in \ref{apx:improvements}. Future developments and prospects will be discussed within the H.E.S.S. Collaboration and the CTA Consortium. Meanwhile, we warmly welcome collaboration with anyone interested in writing up follow-up proposals for deeper MW observations and/or in carrying out dedicated analyses with the existing MW archival data towards the most valuable objects presented here, in particular those not catalogued in the HGPS.

%------------acknowledgements ---------------------------------
%---------------------------------------------------------------------------
\section*{Acknowledgements}
\small
The authors acknowledge the Centre National de la Recherche Scientifique and the Institut National de Physique
Nucl\'eaire et de Physique des Particules in France,  and the Max-Planck-Institut f\"ur Kernphysik in Germany for their support to this project. 
We thank the anonymous referee for his/her valuable comments.
\normalsize

%%%------------bibliography ---------------------------------
\bibliographystyle{elsarticle-harv}
\bibliography{biblioG}

%------------Annex ---------------------------------
\appendix

\section{Construction of the simulated datasets}
\label{apx:Sim_data}

\subsection{Bootstrap simulations}

We performed a bootstrap of the sources catalogued in the HGPS survey. We generated 10 sky realisations, each one including 100 sources. Their latitude, longitude and radius are randomly and independently  picked up among the catalogued value. The spatial distribution of the sources is preserved if the longitude and latitude distributions are independent. The radius is given either by the source size column when available or by the $R_{70}$ column otherwise. For a given radius, we use the same flux (column \textit{Flux\_map} if available) and the same spatial model than catalogued so the surface brightness and the source population distributions are preserved. However for sources  composed of multiple Gaussian components we generated only one Gaussian corresponding to the averaged source given in the HGPS because we do not have the information on the flux of its Gaussian components. This is not an issue as the random re-sampling of the sources will naturally create multiple nested sources. For simplicity the shells are generated with an unique width fixed to 20\% of the inner radius. We simulated shell-like, Gaussian-like and point-like sources using the spatial models implemented in the \textit{Gammapy v0.1} Python package \citep{2017ICRC...35..766D}. The flux map of each source is given by the spatial model (whose integral equals one) times the integrated flux from the bootstrapped catalogue.

\subsection{Toy-model for background emission and exposure maps}

At VHE energies, the \g-ray flux from resolved sources dominates the total emission observed but a large scale diffuse emission is still detected \citep{2008ApJ...688.1078A,2014PhRvD..90l2007A}.  The nature of this emission is assumed to be a mixture of \g rays from unresolved sources and \g rays of interstellar origin. The latter are indirectly produced by the interaction of cosmic ray protons with the interstellar gas (production and decay of neutral pions), and by energy losses of cosmic ray electrons and positrons via Bremsstrahlung and inverse-Compton scattering on radiation fields \citep[see][and reference therein]{2012ApJ...750....3A,2014PhRvD..90l2007A,2016ApJS..223...26A}.

We did not intend to produce a realistic model for the various components of the background emission, instead we used a toy-model that simply mimic a fake contribution from the interstellar \g rays and the instrumental noise. The flux of the toy-model, $F_{\rm Toy-BG}$, is define as   
\begin{equation}
F_{\rm Toy-BG}=F_{GC}  \, (C_{iso}+\, C_{ism}\tau_{353}^{norm})
\end{equation}
with $F_{GC}=1.65\times10^{-12}/[\pi (R_{corr}/d_{pixel})^2]$ cm$^{-2}$ s$^{-1}$ pixel$^{-1}$, the HGPS flux at the Galactic center weighted by the surface ratio between the correlation kernel of the HGPS flux map and the pixel size $d_{pixel}$.

In order to imprint a shape to the diffuse Galactic emission from the interstellar medium, we used the dust optical depth at 353 GHz \citep[$\tau_{353}$ from][]{2016A&A...586A.132P}, normalized by its maximum value, $\tau_{353}^{norm}=\tau_{353}/\tau_{353}^{max}$. Note that the emission from large dust grains and the \g-ray emission from CR interaction with molecular clouds are both dependent on the gas column density so they share many structural similarities. However the dust optical depth per gas nucleon also depends on the grain properties that change across the gas phases and the Galaxy while the \g-ray emissivity per gas nucleon independently varies with energy and Galactocentric radius. So formally the dust emission cannot be used to linearly trace the gas column density and precisely model the shape of the \g-ray emission of interstellar origin. Nevertheless we use this template parametrisation for simplicity as we only need to test how the source reconstruction is affected by an increase in the diffuse background level. To do so we arbitrarily tweaked the contrast in flux between the source and the diffuse background with one parameter. The contrast parameter $C_{ism}$ has been set to 0.1 in the minimal case, which is about the ratio in flux on the HGPS map between the pixels at the Galactic center and the surrounding ones toward the central molecular zone clouds.  We also considered a case where $C_{ism}$=0.3 for a stronger diffuse background emission.

The isotropic term, $C_{iso}$, implicitly assumes that the instrumental noise (hadron background) is proportional to the exposure which is not exactly the case. With such an approximation, we can write the OFF counts as:
\begin{equation}
N_{\rm OFF} /k_{pixels} = B \, t \backsimeq Expo \times (C_{iso}F_{GC})
\label{eq:Noff}
\end{equation}
with $B$ the background count rate, $t$ the observation time and $Expo=A t$ the exposure for the instrument acceptance $A$. As $N_{\rm OFF}$ is integrated within a radius R$_{spec}$, we have to normalize by the surface ratio between the integration radius and the pixel size, \mbox{$k_{pixels}=\pi (R_{spec}/d_{pixel})^2$}.

\begin{figure}
\centering 
\includegraphics[width=\hsize]{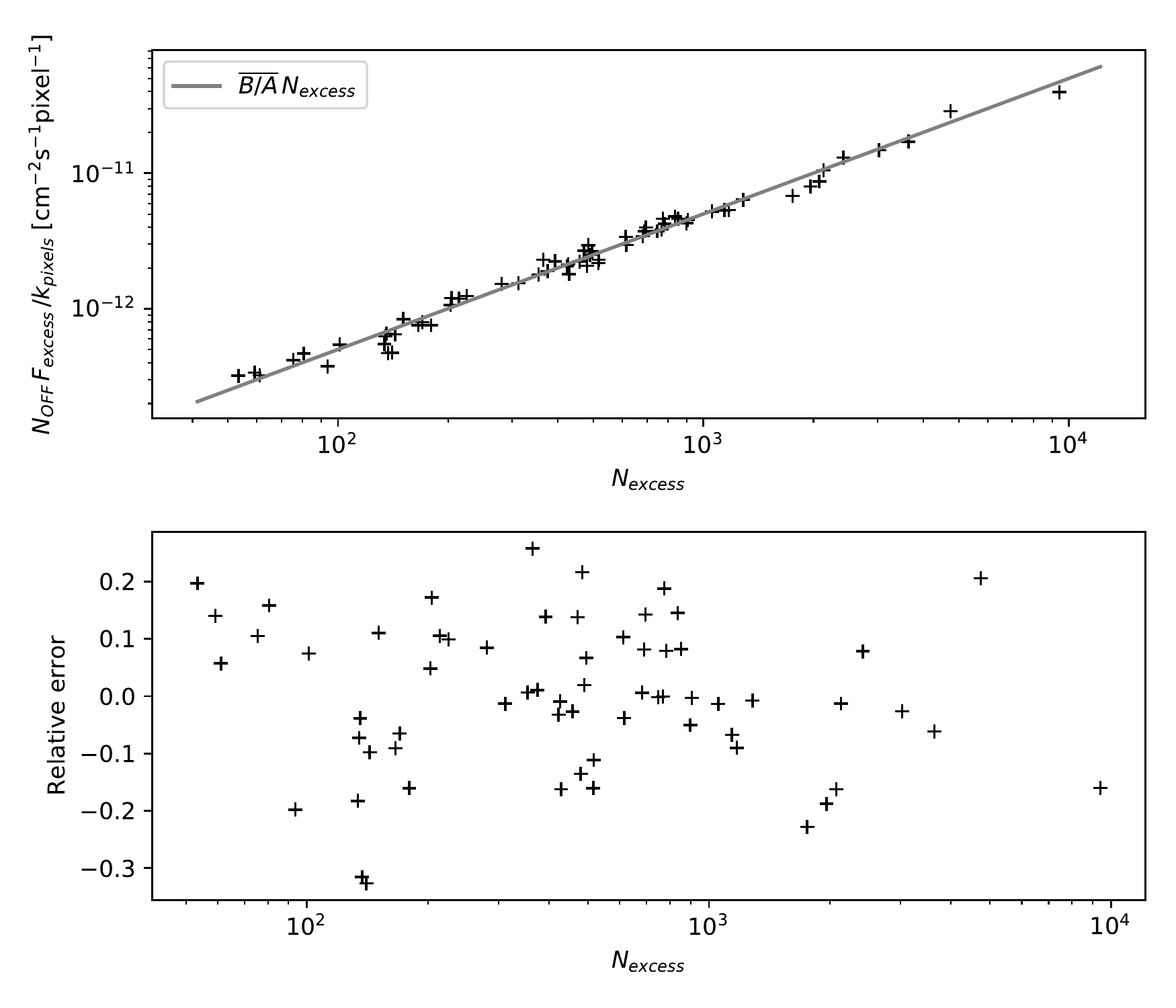}  
\caption{Average background noise parameter -- Top: Estimation of B/A according to Eq. \ref{eq:BsurA} using the $N_{excess}$, $F_{excess}$ and $N_{\rm OFF}$ of the sources fitted in the HGPS catalog. The grey line correspond to a slope equal to the average $B/A$ ratio per pixel. Bottom: Relative error on the average $B/A$ value.}
\label{fig:BsurA}
\vspace{1cm}
\centering 
\includegraphics[width=\hsize]{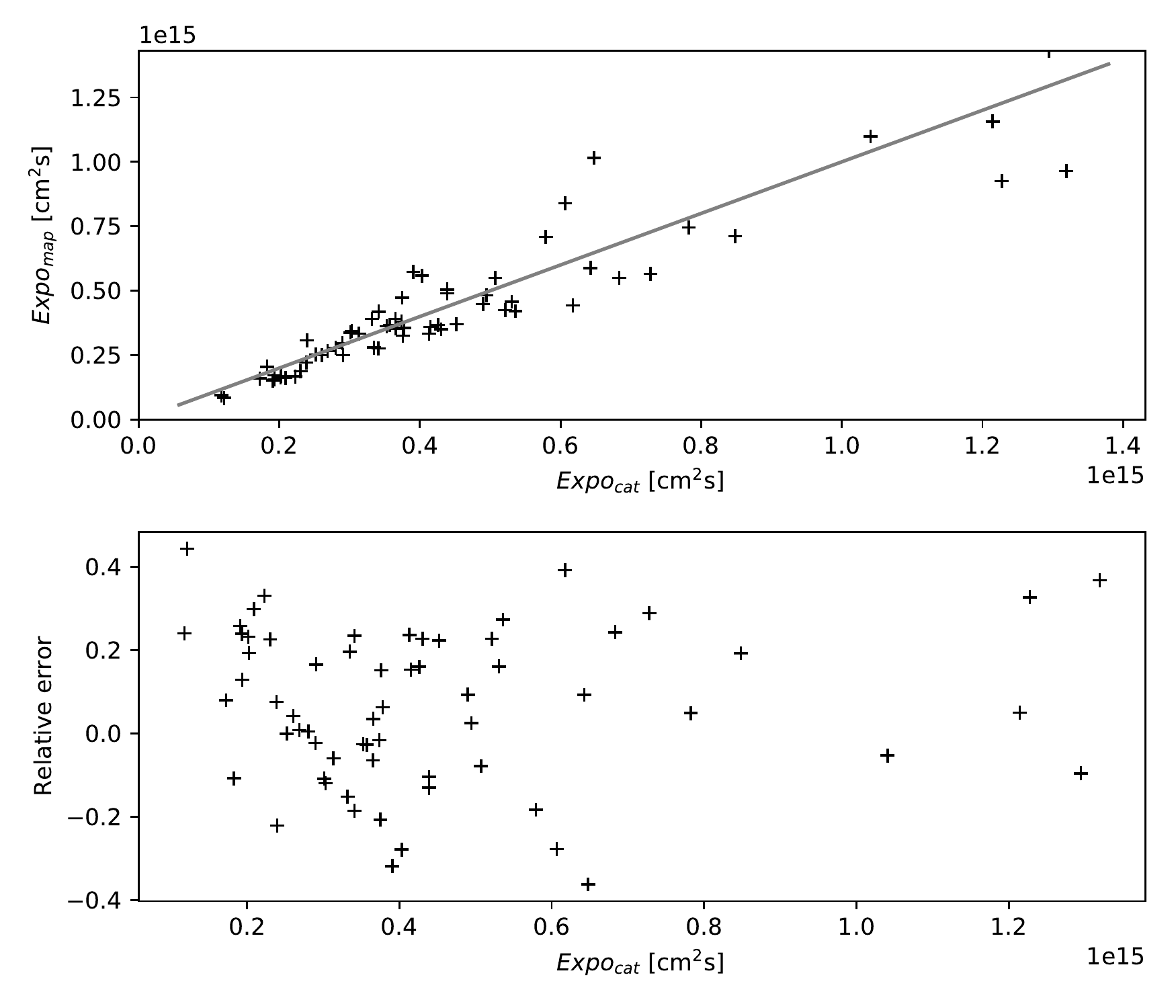}  
\caption{Comparison of exposures between catalogue and simulations -- Top: Exposure estimated for each source in the catalogue using Eq. \ref{eq:ExpoLim}, $Expo_{cat}$~\textit{versus} the exposure from the simulated map at each source location, $Expo_{map}$. The grey line corresponds to a one-to-one correlation. Bottom: Relative error on the simulated exposure as $(Expo_{cat}-Expo_{map})/Expo_{map}$.}
\label{fig:Expo_test}
\end{figure}

Then the normalisation factor $C_{iso}$ is proportional to the ratio $B/A$, that can be estimated using the information on the sources fitted in the HGPS catalogue.
Indeed the excess counts are expressed as:
\begin{equation}
N_{excess}= A \, t \, F_{excess} =Expo \, F_{excess}
\label{eq:Nexcess}
\end{equation}
so combining the two previous equations we have:
\begin{equation}
\frac{B}{A}=\frac{N_{\rm OFF}F_{excess}}{k_{pixels} N_{excess}} \backsimeq C_{iso}F_{GC}
\label{eq:BsurA}
\end{equation}
with $N_{excess}$, $F_{excess}$ and $N_{\rm OFF}$ given in the HGPS catalog as the columns named \textit{Excess\_RSpec\_Model}, \textit{Flux\_Map\_RSpec\_Total}, and \textit{Background\_RSpec}, respectively.
We get an average $B/A$ ratio per pixel: \mbox{$\overline{B/A}=5.01\times 10^{-15}$ cm$^{-2}$s$^{-1}$pixel$^{-1}$} (see Fig. \ref{fig:BsurA}). Then we have the normalization factor \mbox{$C_{iso}\simeq0.24$}.

%The simulated exposure is derived from the sensitivity maps of the HGPS survey, as follow.  
In the limit case where $N_{\rm OFF} \gg N_{excess}$, the significance of the excess can be expressed as
 $\sigma_{excess}=N_{excess}/\sqrt{N_{\rm OFF}}$.
By combining this relation with the Eq. \ref{eq:Nexcess} into Eq. \ref{eq:BsurA}  we can write:
\begin{equation}
\rm{Expo_{lim}}=\pi\left(\frac{R_{spec}}{d_{pixel}}\right)^2\overline{\left(\frac{B}{A}\right)}\left(\frac{\sigma_{excess}}{F_{excess}}\right)^2
\label{eq:ExpoLim}
\end{equation}
Note that we can also derive an average $B/A$ from this equation and the value found is consistent with the previous one even if we approximate the significance.

The HGPS sensitivity map is defined as the minimal flux in cm$^{-2}$s$^{-1}$ within a correlation radius $R_{corr}$ required to detect a source with a significance of 5. Thus, according to Eq. \ref{eq:ExpoLim}, we can define the simulated exposure map in cm$^{2}$s as :
\begin{equation}
\rm{Expo_{lim}}=\pi \left(\frac{R_{corr}}{d_{pixel}}\right)^2\overline{\left(\frac{B}{A}\right)}\left(\frac{5\sigma}{\rm Sensitivity}\right)^2
\end{equation}
The same exposure map has been used for all the simulations. In Fig. \ref{fig:Expo_test} we checked that the exposure estimated for each source in the catalogue using Eq. \ref{eq:ExpoLim} and the simulated exposure map at each source location are consistent.
 
\subsection{Count and significance maps}

The simulated flux maps of the sources and of the background have been multiplied by the exposure map and convolved by a Gaussian kernel of $0.06$\degree~in order to mimic the instrument point-spread-function. We then applied Poisson noise to get the simulated counts. 

The significance maps have been derived using the same formalism as that of the HGPS survey. The excess counts $N_{excess}$ within an ON region are defined as
\begin{equation}
N_{excess} = N_{\rm ON} - \alpha N_{\rm OFF}
\end{equation}
with $N_{\rm ON}$ the counts in the ON region, $N_{\rm OFF}$ the counts in the OFF region, and $\alpha$ the background normalization factor defined as the ratio of the integral acceptances in the ON/OFF regions: $\alpha=\xi_{\rm ON}/\xi_{\rm OFF}$.
The significance has then been calculated using Eq. (17) from \cite{1983ApJ...272..317L}:
\begin{align}
S=\sqrt{2}&\left\lbrace  N_{\rm ON} \ln \left[  \frac{1+\alpha}{\alpha} \left(\frac{N_{\rm ON}}{N_{\rm ON}+N_{\rm OFF}}\right)\right] + \right. \\ \notag
&\ \left. N_{\rm OFF} \ln \left[  (1+\alpha) \left(\frac{N_{\rm OFF}}{N_{\rm ON}+N_{\rm OFF}}\right)\right] \right\rbrace  ^{0.5}
\end{align}
The ON region is defined as a disk of radius $R_{corr}$ (correlation radius). We produced maps with $R_{corr}=0.1 \degree$ and $0.2 \degree$.
The OFF counts are determined either from the expected counts of the simulated background or from the adaptive ring background method.   
In the first case the OFF region is the same as the ON region since we have a perfect knowledge of the OFF counts.
In the second case we proceed in a similar way as that developed in the HGPS. For each pixel, the OFF region is defined as a concentric ring with inner radius of 2 times $R_{corr}$ and a variable outer radius chosen to minimize the difference to a target $\alpha$ value fixed to 0.02. By comparison the default geometric $\alpha$ of the HGPS analysis is about 0.01 (surface ratio between ON/OFF regions)
%\footnote{HGPS default geometric $\alpha$:
%\begin{align*}
%\alpha_{geom}&=R_{\rm ON}^2/(R_{\rm OFF,out}^2-R_{\rm OFF,in}^2) \\
%			&=0.1^2/(1.14^2-0.7^2) \\
%			&\simeq 0.012
%\end{align*}
%}.
%\mr{(Is this footnote necessary?)}
After producing a first significance map we define an exclusion region in order to mask the significant pixels from the OFF region. Then the background is re-evaluated in each pixel using the masked adaptive ring, and the significance map is updated. This iterative procedure is repeated three times (after that the limited change in the exclusion mask no longer affects the final results). 

The exclusion mask is derived by hysteresis thresholding on the significance maps with a high threshold of $4.5\sigma$  and a low threshold of $1\sigma$. Note that this step differs from the HGPS procedure where a hard threshold at $5\sigma$ is used and the mask is then uniformly extended by $0.3\degree$ beyond the $5\sigma$ contours. The mask obtained by hysteresis thresholding better preserves the structure of the excess.

\begin{figure*}[!p]
\centering 
\includegraphics[width=\hsize]{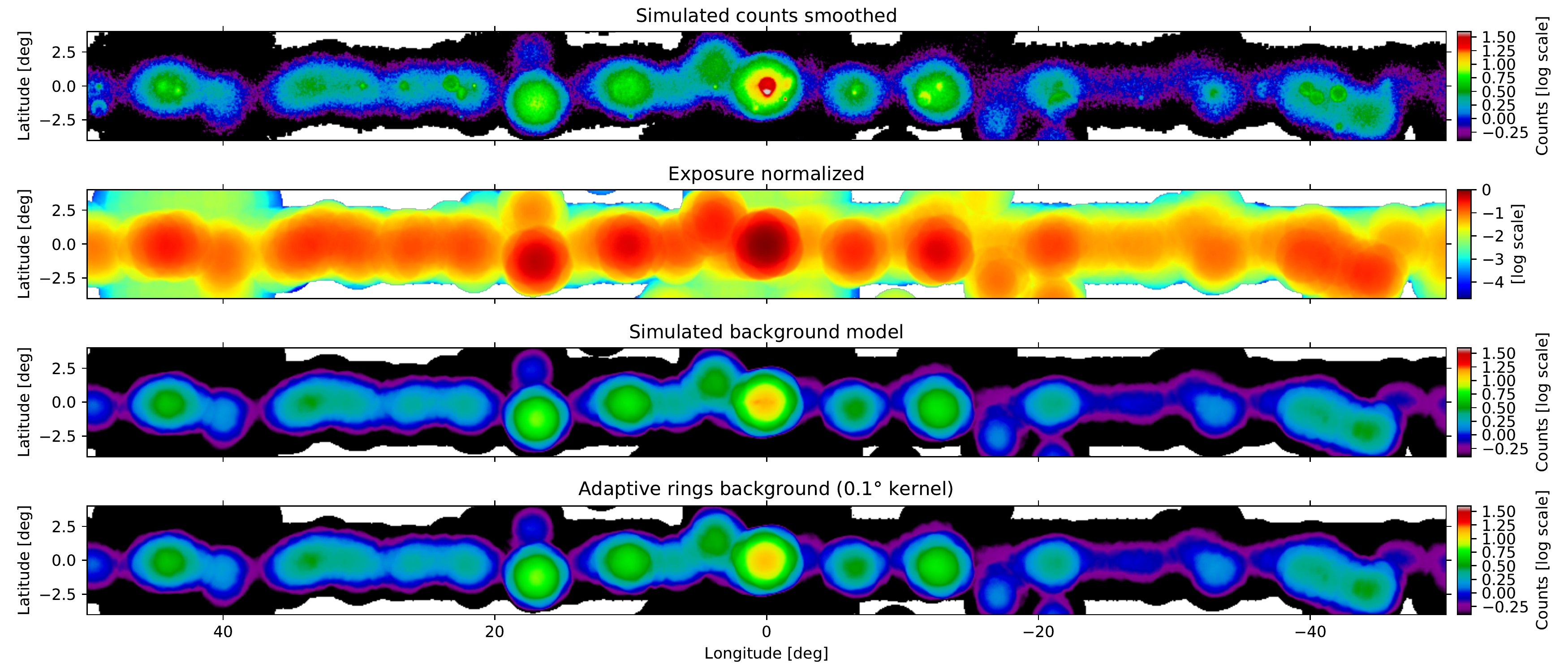}  
\caption{Simulated sky maps ( -- First row: Count map. Second row: Exposure map normalized by its maximum value. Third row: Toy-model for the background emission used in the simulation. Fourth row: Background emission derived using the adaptive ring background method (R$_{corr}$ = 0.1\degree). The example displayed correspond to the first simulated sky in Fig. \ref{fig:Signi_sample} (second row)}
\label{fig:Sim_data}
\end{figure*}

\begin{figure*}[!p]
\centering 
\includegraphics[width=\hsize]{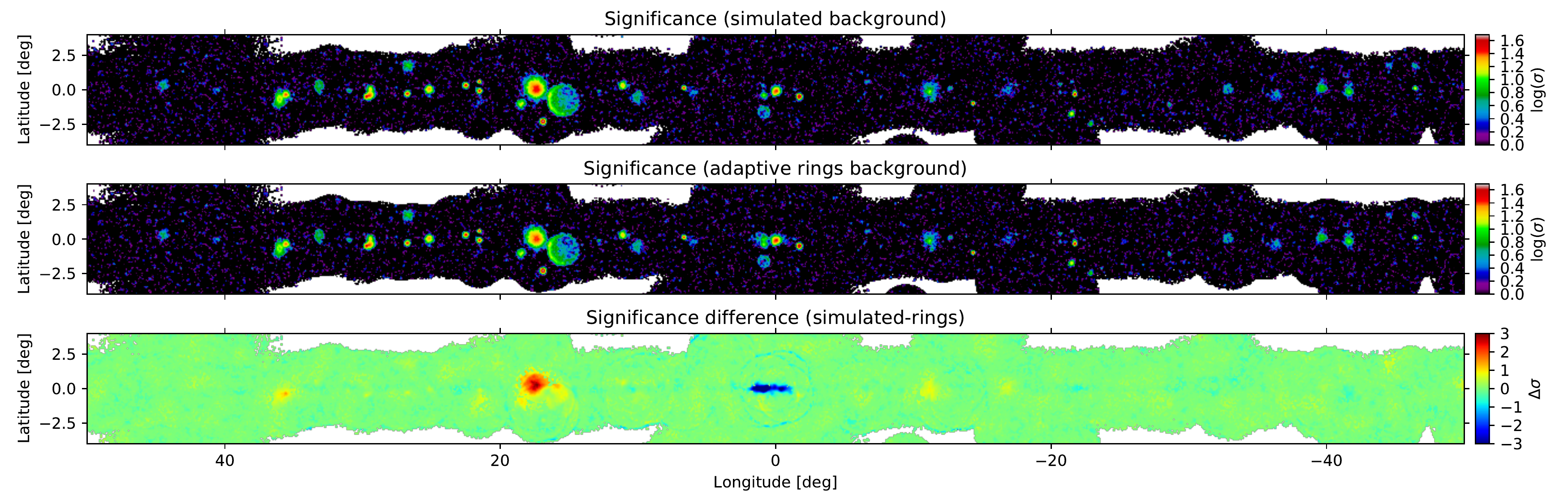}  
\caption{Simulated significance maps -- First row: Significance map derived using the exact background model (with C$_{ism}$=0.1, see \ref{apx:Sim_data}); Second row: significance map derived using the adaptive ring background method (see Fig.\ref{fig:Sim_data}); Third row: Difference in significance between the two maps. Note that significance maps are produced for a correlation radius \mbox{R$_{corr}$ = 0.1\degree}.}
\label{fig:Sim_Signi}
\end{figure*}

\begin{figure*}[!p]
\centering 
\includegraphics[width=\hsize]{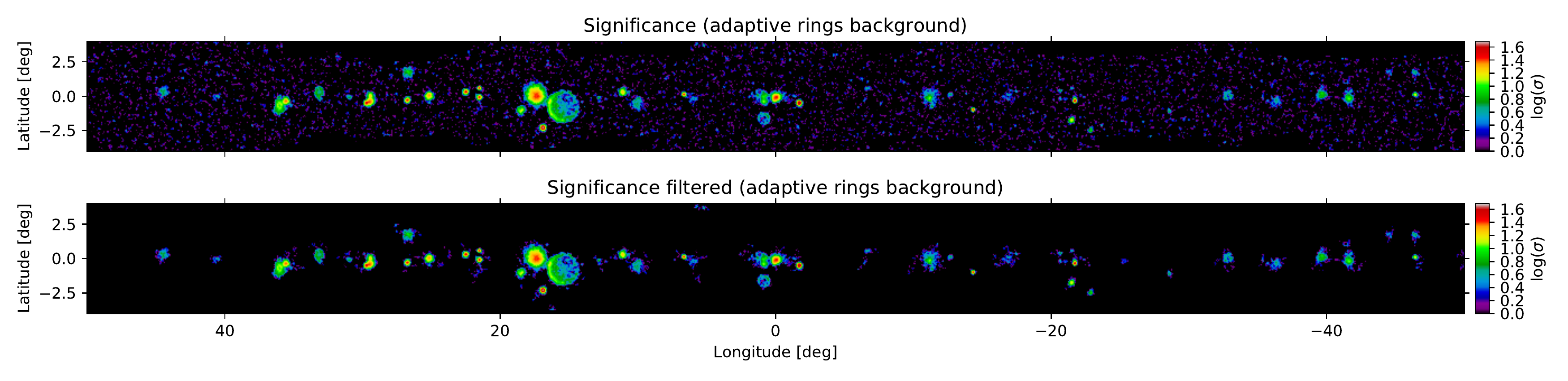}  
\caption{Significance map filtering -- Top: Significance map derived using the adaptive ring background method ($R_{corr} = 0.1\degree$, with C$_{ism}$=0.1). Bottom: Same after filtering by hysteresis thresholding (using $\sigma_{high}=4.5$, $\sigma_{low}=1$).}
\label{fig:Sim_SigniFIL}
\end{figure*}

In Fig. \ref{fig:Sim_data} we show the counts, exposure and background maps for one of the simulated skies.
In Fig. \ref{fig:Sim_Signi} simulated significance maps derived from the exact background and through the adaptive ring background method, together with their difference, are presented. In the edge detection procedure we filter these significance maps using the information of the maps with $R_{corr} = 0.1\degree$ and $R_{corr} = 0.2\degree$ by combining their respective hysteresis mask with an \textit{or} condition. An example of filtered significance map for a simulated sky is shown in Fig. \ref{fig:Sim_SigniFIL}.

\begin{figure*}[!p]
\centering 
\includegraphics[width=\hsize]{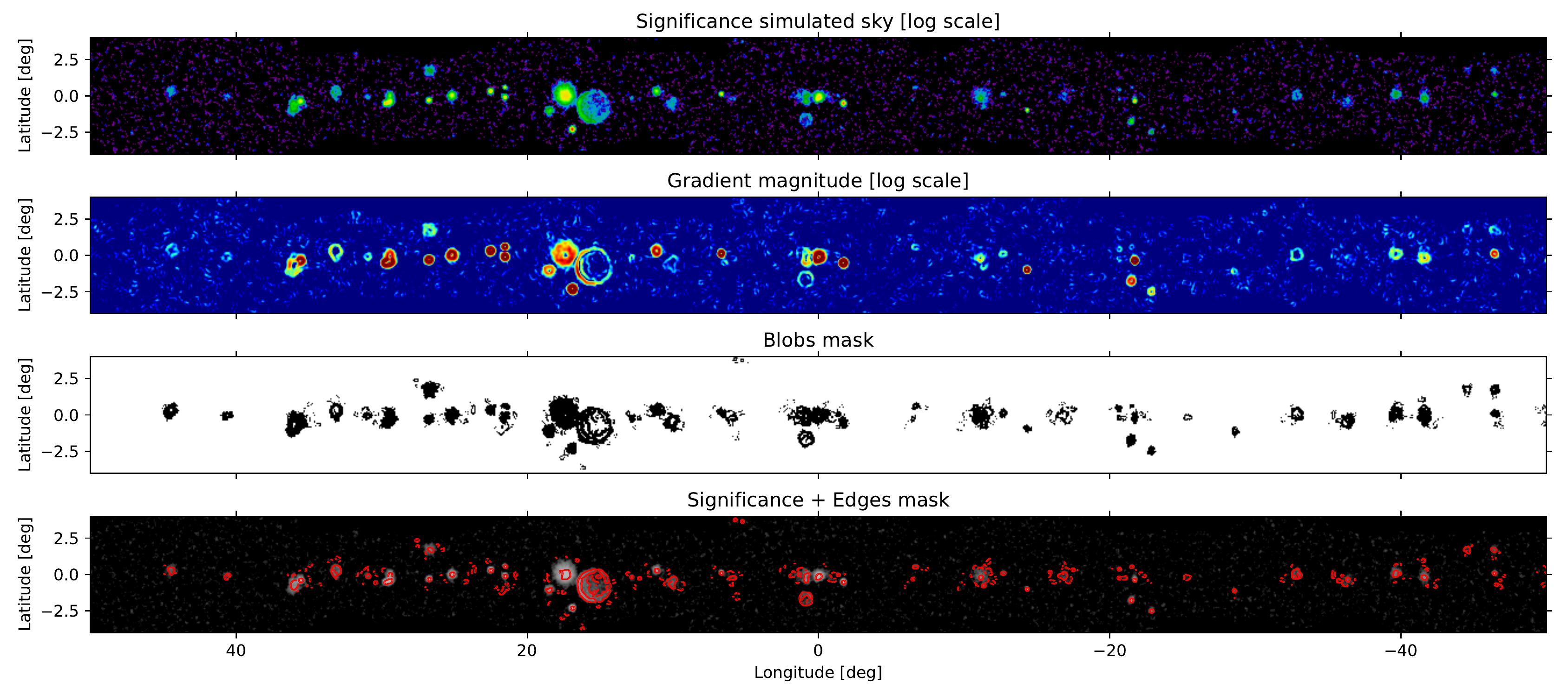}  
\caption{Edge detection on simulated significance map -- First row: Significance map derived using the Adaptive Ring Background method as in the HGPS (R$_{corr}$ = 0.1\degree). Second row: Gradient of the significance map derived using Sobel operator. Third row: Blob mask given by thresholding the significance and gradient values (as described in Sect.~\ref{sec:prepro}). Fourth row: Significance map at R$_{corr}$ = 0.1\degree~(same as first row but in grey-scale) with the Edge mask obtained after applying the Canny filter overlaid in red.}
\label{fig:Sim_preprocess}
\vspace{2cm}
\centering 
\includegraphics[width=\hsize]{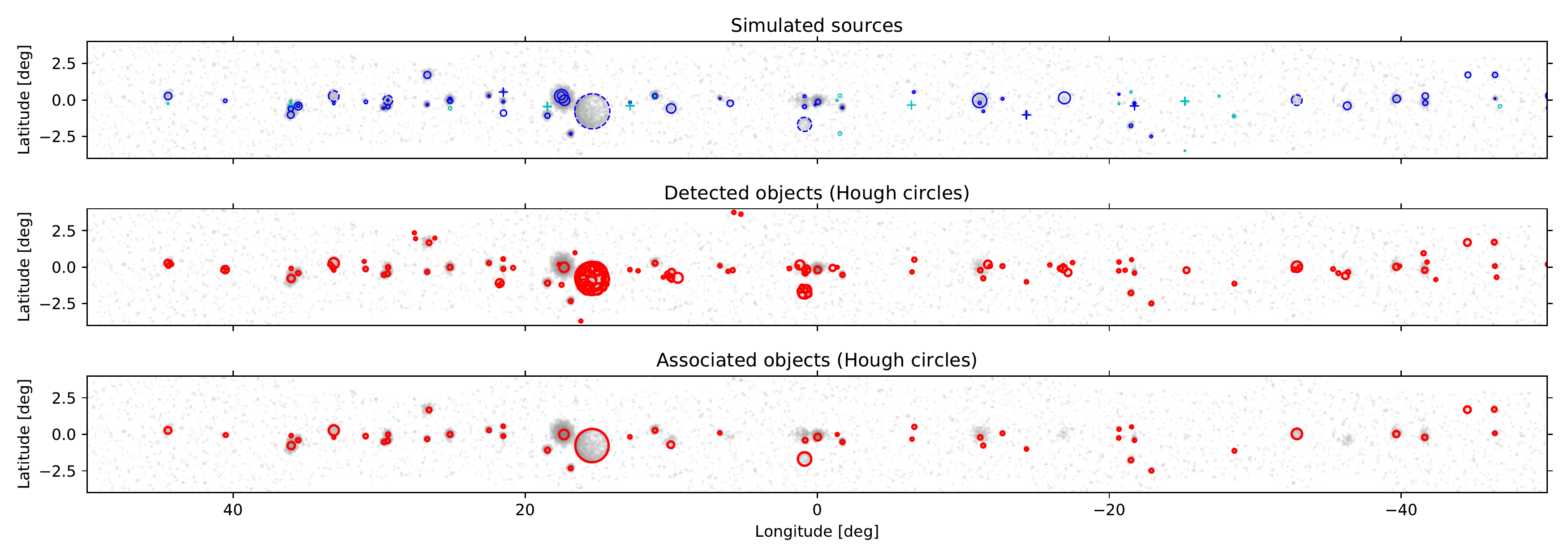}  
\caption{Comparison of detected objects with simulated sources -- In all panels the underlying map is the same significance as that shown in Fig. \ref{fig:Sim_preprocess},  derived using the Adaptive Ring Background method as in the HGPS (R$_{corr}$ = 0.1\degree). First row: Simulated sources, the dotted circles correspond to the outer radii of radial shells, the plain circles correspond to the sigma of radial Gaussians, point sources are shown as crosses. Light and dark blue colors correspond to sources with an expected TS value below and above 30, respectively. Second row: objects detected in Hough space. Third row: Detected objects associated to simulated sources  using criterion on inter-center distance and surface overlap as described in Sect.~\ref{sec:xmatch}. Note that we test for associations regardless of the expected TS values.}
\label{fig:Sim_sources_comp}
\end{figure*}

\section{Possible technical improvements}
\label{apx:improvements}

\vspace*{0.03cm} \textbf{\noindent On the background removal:}\\
Using the gradient information to extract only sparse structural information and detect only sharp edges in the data allow us in principle to filter most of the diffuse structures associated to the interstellar background. It should be further tested on the on-going simulations of the Galactic Plane Survey which will be conducted by the next generation of imaging atmospheric Cherenkov telescopes (IACTs), namely the Cherenkov Telescope Array (CTA) \citep{2011ExA....32..193A,2017arXiv170907997C}. CTA will probe lower energies with a higher sensitivity and better angular resolution so that the contribution of diffuse Galactic emission will be larger and hence, more refinement may be necessary to filter the diffuse background. The filters applied to the edge mask can enhance the response of the algorithm to sources against background.\\

\textbf{\noindent On the object detection:}\\
In complement to object detection based on the Hough transform, we could use the Wavelet transform. Both Wavelet-based and Hough-based detections can be used to generate a list of seeds in a short amount of time that can be further tested in dedicated analyses or catalogue pipelines using the template-fitting approach. According to our tests, a good agreement in position and radius is found between the two detection methods for small-size objects resolved in Hough space. The wavelets-based detection presents a clear advantage for dissociating point-like/small-size objects in close vicinity. The Hough-based detection presents a huge advantage for extended objects, in particular SNR-like objects for which the shell is not entirely above the noise level. It would be interesting to further test the complementarity of these two methods in the framework of the CTA simulations, in order to help in the preparation of the future catalogue pipeline. \\

\textbf{\noindent On the object classification and identification:}\\
We have shown that the different types of well-identified sources can be separated within a set of morphological parameters using simple linear classification rules. More sophisticated machine-learning techniques like support vector machines \citep{Cortes1995} could be used to define the classification rules automatically. In our study, we have used the well-identified HGPS sources to define a training sample and find new source candidates that share similar characteristics. In a more general way we could use unsupervised machine-learning techniques as clustering algorithms to identify different classes of objects and identify them to known source populations only a posteriori. This could help dissociate sources from spurious detections or find new types of VHE-emitting sources (not represented in any training sample). Of course the more object classes we want to separate, the higher the dimension of the parameter space should be. For instance, we could improve the classification of the different source types by including their spectral index and their variability.\\

\textbf{\noindent On the multi-wavelength associations:}\\
In order to firmly confirm the nature of these objects, dedicated MW analyses, including spectral fitting, are necessary. We postpone this work for future follow-up studies. As MW catalogues are not exhaustive, deep mining into MW archival data is essential to find potentially relevant counterparts that have remained below the detection threshold of catalogues. Given the number of available surveys from radio to \g-rays, collecting all these data is a complicated and time-consuming task. Moreover cross-matching all these surveys is far from being straightforward given the large diversity in the instrumental performances and characteristics in terms of calibration, resolution, field-of-view and so on. In order to address this important issue, a new automated algorithm is being developed by J. Devin and M. Renaud. In the future, it could be used in order to further look for MW coincidences toward the most interesting source candidates flagged by the morphological classifier, especially those which do not show any catalogued counterpart.

%\clearpage
\section{Additional figures and tables}

\begin{figure}[!b]
\centering
\includegraphics[scale=0.48]{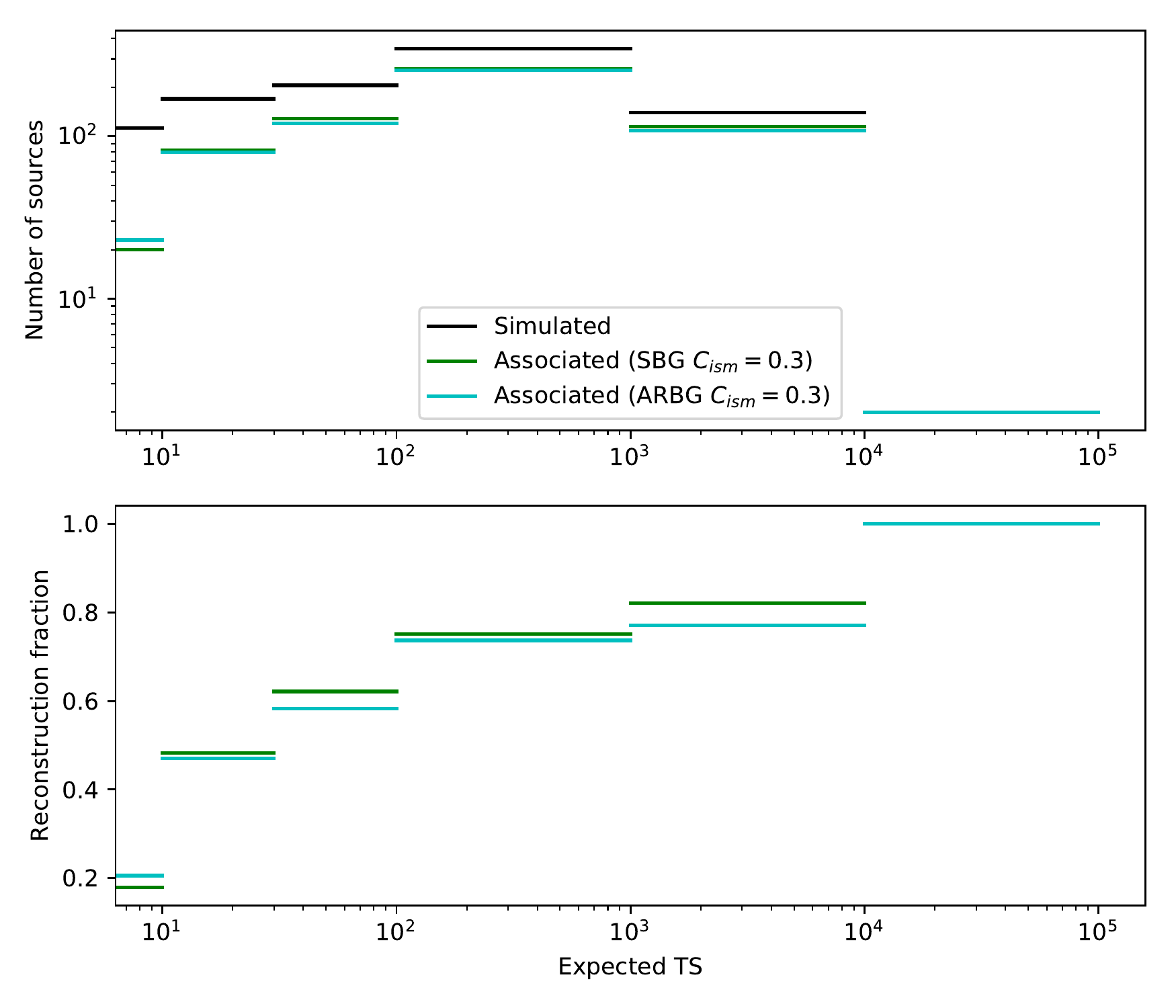}  
\caption{Reconstruction fraction with TS -- Same as Fig. \ref{fig:Sim_detfracTS} but for a higher background contagion level ($C_{ism}=0.3$).  Top: Histogram of the number of sources within bins in expected TS values. Bottom: Fraction of simulated sources associated to a detected object in the same bins. Green and cyan bars correspond to the objects detected as Hough circles on the significance map derived using simulated background (SBG) and the adaptive ring background method (ARBG), respectively.}
\label{fig:Sim_detfracTS3}
\includegraphics[scale=0.48]{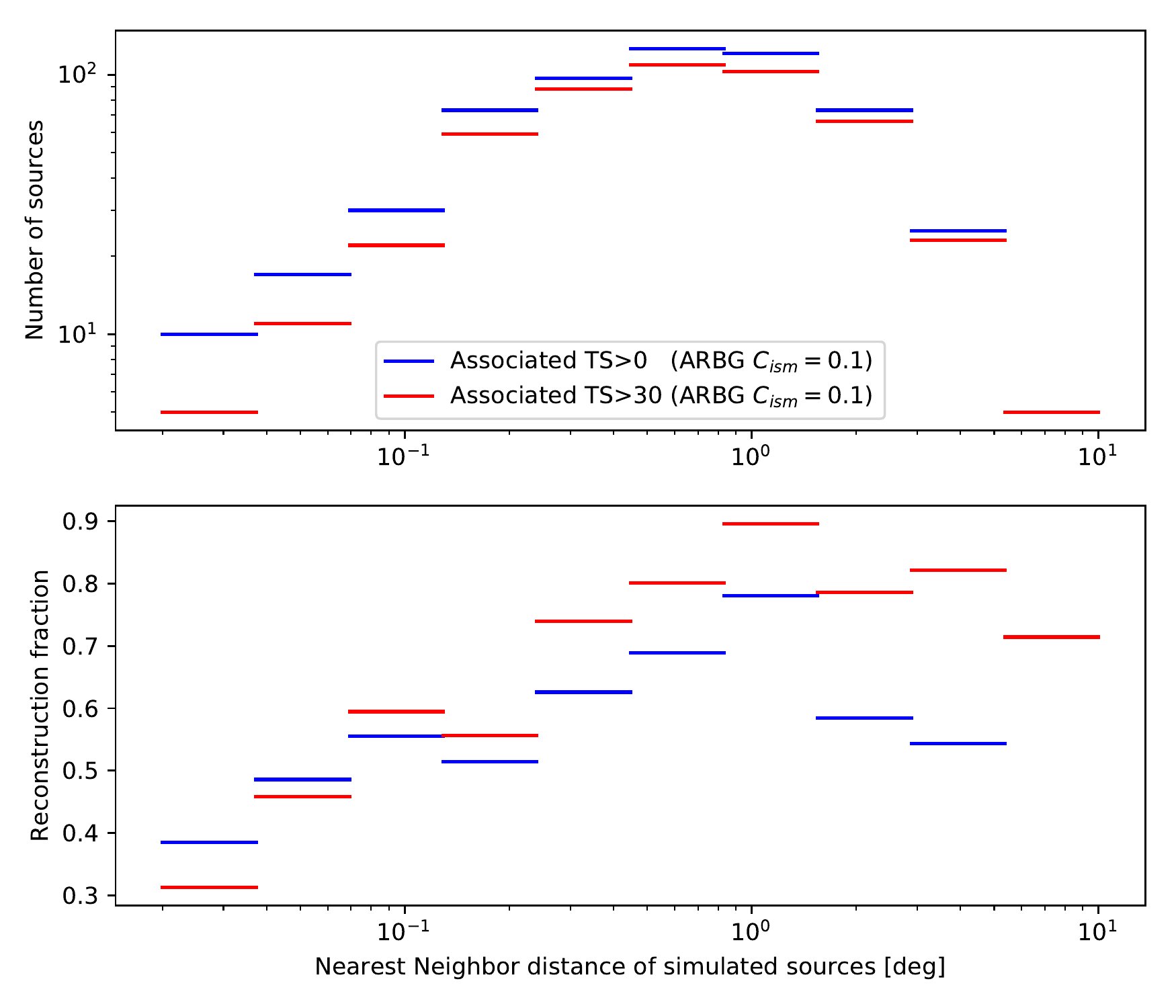}  
\caption{Reconstruction fraction with the nearest neighbor distance d$_{nn}$ -- Top: Histogram of the number of sources within bins in d$_{nn}$. Bottom: Fraction of simulated sources associated to a detected object in the same bins. Blue and red bars correspond to the reconstruction fraction of sources with an expected TS larger than 0 and 30, respectively. In both cases we derive significance maps using the adaptive ring background method (ARBG) for the simulated skies with the minimal background level ($C_{ism}=0.1$).}
\label{fig:Sim_detfracDNNTS}
\end{figure}

\label{apx:add}
\renewcommand{\arraystretch}{0.7}
\setlength{\tabcolsep}{0.1cm}
\begin{table*}[h]
\caption{Description of the \textit{FITS} tables columns}
\centering
\resizebox{\textwidth}{!}{%
\begin{tabular}{c|c}
\hline
\hline
Column name & Description \\
\hline
& \\
Name& Object identifier by reference number\\
& \\
GLON& Galactic longitude of the object center in degree\\
& \\
GLAT& Galactic latitude of the object center in degree\\
& \\
Radius& Radius of the object in degree\\
& \\
Circularity& Fraction of circle circumference fraction that have triggered the detection in Hough space\\
& \\
SigCenter& Significance at the central pixel of the object (in the map with R$_{corr}$ = 0.1$\degree$)\\
& \\
NestFlag& Nesting flag: indicate the level of overlapping between obects based on inter-center distances (see Sect. \ref{sec:params})\\
& \\
TypeMorph& Morphological flag: peak, flat or cavity (see Sect. \ref{sec:params})\\
& \\
PearsonR&  Pearson correlation coefficient  of a 5-point radial profile in flux (see Sect. \ref{sec:params})\\
& \\
\multirow{2}{*}{HGPS\_Region} & Name of the HGPS region emcompassing the object (see Sect. \ref{sec:xmatch})\\
  & Regions are defined by the extraction radius of the source if available or its size otherwise \\
& \\
HGPS\_Object& Name of the HGPS source (see Sect. \ref{sec:xmatch})\\
& \\
\multirow{2}{*}{HGPS\_Otype} & Type of source as given by the HGPS catalog\\
 & differentiation between SNR and snr-mc based on the gamma-cat (see Sect. \ref{sec:HGPScomp})\\
& \\
HGPS\_Radius& Radius of the HGPS source \\
& \\
HGPS\_Dist& Inter-center distance between the object and the HGPS source associated \\
& \\
HGPS\_SFoverlap& Surface overlap fraction between the object and the HGPS source (see Sect. \ref{sec:xmatch})\\
& \\
HGPSG\_Object& HGPS Gaussian component(see Sect. \ref{sec:xmatch})\\
& \\
HGPSG\_Radius& Radius of the HGPS Gaussian Component ($\sigma$-width)\\
& \\
HGPSG\_Otype& Component class of the HGPS Gaussian \\
& \\
HGPSG\_Dist& Inter-center distance between the object and the HGPS Gaussian component associated \\
& \\
HGPSG\_SFoverlap& Surface overlap fraction between the object and the HGPS Gaussian component (see Sect. \ref{sec:xmatch})\\
& \\
SNRCAT\_Object& Name of the radio source in the SNRcat (see Sect. \ref{sec:cats}) \\
& \\
SNRCAT\_Otype& Type of the SNRcat source (Type column from the SNRcat reference in the HGPS catalog) \\
& \\
SNRCAT\_Radius& Radius of the SNRcat source (RADIO\_RADIUS column from the SNRcat reference in the SC1 catalog) \\
& \\
SNRCAT\_Dist& Inter-center distance between the object and the SNRcat source associated \\
& \\
SNRCAT\_SFoverlap& Surface overlap fraction between the object and the SNRcat source (see in Sect. \ref{sec:xmatch})\\
& \\
SC1\_Object& Name of the \textit{Fermi}-LAT source from the SC1 catalog (see Sect. \ref{sec:cats}) \\
& \\
SC1\_Otype& Type of the SC1 source (Classification column of the original catalog) \\
& \\
SC1\_Radius& Radius of the SC1 source  \\ 
& \\
SC1\_Dist& Inter-center distance between the object and the SC1 source associated \\
& \\
SC1\_SFoverlap& Surface overlap fraction between the object and the SC1 source (see Sect. \ref{sec:xmatch})\\
& \\
FL8Y\_Object& Name of the  \textit{Fermi}-LAT source from the FL8Y list (see Sect. \ref{sec:cats}) \\
& \\
FL8Y\_Otype& Type of the  FL8Y source (CLASS column of the original catalog) \\
& \\
FL8Y\_Dist& Inter-center distance between the object and the FL8Y source associated \\
& \\
FL8Y\_SFoverlap& Surface overlap fraction between the object and the FL8Y source (see Sect. \ref{sec:xmatch})\\
 \hline
 \hline
\end{tabular}}
\label{tab:content}
\end{table*}
\clearpage
\renewcommand{\arraystretch}{0.7}
\setlength{\tabcolsep}{0.1cm}
\begin{table*}[h]
\caption{List of candidates based on high-circularity filter and MW-coincidence search (see Sect.~\ref{sec:PScand})}
\centering
\resizebox{\textwidth}{!}{%
\begin{tabular}{c|ccccc}
\hline
\hline
& \\
Name& HC\_43 & HC\_56 & HC\_57 & HC\_115 & HC\_258\\
& \\
GLON& -93.01 &-68.97 & -68.81 & -39.41 & 0.63\\
& \\
GLAT& -0.97 & -0.01 & -0.53 & -0.01 &2.27\\
& \\
Radius& 0.16 &0.14 & 0.14 & 0.11 &0.12\\
& \\
Circularity& 0.98 & 1 & 1 & 1 & 1\\
& \\
SigCenter& 10.51 & 3.19 & 4.40 & 4.09 & 4.19\\
& \\
NestFlag& 3 & 0 & 0 & 0 & 0 \\
& \\
TypeMorph& Peak & Peak & Peak & Peak & Peak \\
& \\
PearsonR& -0.99 & -0.99 & -0.98 & -0.95 & -0.97\\
& \\
HGPS\_Region& & & & &\\
& \\
HGPS\_Object& & & & &\\
& \\
HGPS\_Otype& & & & &\\
& \\
HGPS\_Radius& & & & & \\
& \\
HGPS\_Dist& & & & & \\
& \\
HGPS\_SFoverlap& & & & &\\
& \\
HGPSG\_Object& & & & &\\
& \\
HGPSG\_Otype& & & & & \\
& \\
HGPSG\_Radius& & & & & \\
& \\
HGPSG\_Dist& & & & & \\
& \\
HGPSG\_SFoverlap& & & & &\\
& \\
SNRCAT\_Object& G267.0-1.0 & G291.0-0.1 & & G308.8-0.1 &\\
& \\
SNRCAT\_Otype& filled-centre & plerionic composite & & plerionic composite & \\
& \\
SNRCAT\_Radius& 0. & 0.12 & & 0.2 & \\
& \\
SNRCAT\_Dist& 0.04 & 0.07 \\
& \\
SNRCAT\_SFoverlap& 0.41 &0.5 & & 0.26 &\\
& \\
SC1\_Object& & SNR291.0-00.1 & & & \\
& \\
SC1\_Otype& & Classified & & &  \\
& \\
SC1\_Radius& & 0. & & &   \\ 
& \\
SC1\_Dist& & 0.11 & & &  \\
& \\
SC1\_SFoverlap& & 0.25 & & & \\
& \\
FL8Y\_Object& & 4FGL J1111.8-6039 & 4FGL J1112.1-6108 & 4FGL J1341.7-6216 & 4FGL J1511.2-5803\\
& \\
FL8Y\_Otype& & PSR & PSR & PSR & \\
& \\
FL8Y\_Dist& & 0.10 & 0.05 & 0.05 & 0.06 \\
& \\
FL8Y\_SFoverlap& & 0.29 & 0.52 & 0.55 & 0.51\\
 \hline
 \hline
\end{tabular}}
\label{tab:COMPcand}
\end{table*}
\clearpage
\renewcommand{\arraystretch}{0.7}
\setlength{\tabcolsep}{0.1cm}
\begin{table*}[h]
\caption{List of the SNR-like candidates (see Sect.~\ref{sec:SNRcand})}
\centering
\resizebox{\textwidth}{!}{%
\begin{tabular}{c|ccccccc}
\hline
\hline
& \\
Name& HC\_73 &HC\_76 & HC\_249 & HC\_282 & HC\_303 & HC\_403 & HC\_450 \\
& \\
GLON& -48.27 & -47.65 & -0.97 & 6.35 & 11.43 & 33.23 & 44.47\\
& \\
GLAT& 0.19 & -0.25 & -0.87 & -0.13 & -0.29 & -0.15 & -0.17\\
& \\
Radius& 0.27 & 0.29 & 0.31 & 0.38 & 0.33 & 0.21 & 0.42\\
& \\
Circularity& 0.56 & 0.55 & 0.61 & 0.55 & 0.72 & 0.64 & 0.8\\
& \\
SigCenter& 0.27 & 1.37 & 3.18 & 0.30 & 5.27 & 2.63 & 3.50\\
& \\
NestFlag& 1 & 1 & 1 & 1 & 1 & 1 & 1\\
& \\
TypeMorph& Cavity & Cavity & Cavity & Cavity & Cavity & Cavity & Cavity\\
& \\
PearsonR&  0.53 & 0.54 & 0.61 & 0.90 & 0.48 & 0.86 & 0.54\\
& \\
HGPS\_Region& & & & & HESS J1809-193 & HESS J1852-000 & HESS J1912+101 \\
& \\
HGPS\_Object& & & & & & HESS J1852-000 & HESS J1912+101\\
& \\
HGPS\_Otype& & & & & & Unid & Unid \\
& \\
HGPS\_Radius& & & & & & 0.28 & 0.49\\
& \\
HGPS\_Dist& & & & & & 0.12 & 0.04\\
& \\
HGPS\_SFoverlap& & & & & & 0.46 & 0.70\\
& \\
HGPSG\_Object& & & HGPSC 051 & & HGPSC 059 & HGPSC 090 & \\
& \\
HGPSG\_Otype& & & Discarded Large & & Source Multi & Source Single &  \\
& \\
HGPSG\_Radius& & & 0.40 & & 0.30 & 0.28 &  \\
& \\
HGPSG\_Dist& & & 0.02 & & 0.01 & 0.12 &  \\
& \\
HGPSG\_SFoverlap& & & 0.57 & & 0.90 & 0.46 & \\
& \\
SNRCAT\_Object& & G312.4-0.4 & G359.0-0.9 & G6.4-0.1 & & &\\
& \\
SNRCAT\_Otype& & shell & shell & shell & & &  \\
& \\
SNRCAT\_Radius& & 0.32 & 0.19 & 0.40 & & & \\
& \\
SNRCAT\_Dist& & 0.15 & 0.06 & 0.1 & & & \\
& \\
SNRCAT\_SFoverlap& & 0.53 & 0.48 & 0.70 & & &\\
& \\
SC1\_Object& SNR311.5-00.3 & & SNR358.5-00.9 & & SNR011.4-00.1 & &  \\
& \\
SC1\_Otype& Not an SNR & & Not an SNR & & Marginally Classified &  & \\
& \\
SC1\_Radius& 0.30 & & 2.09 & & 0. & &  \\ 
& \\
SC1\_Dist& 0.04 & & 0.13 & & 0.17 & &  \\
& \\
SC1\_SFoverlap& 0.74 & & 0.02 & & 0.09 & &  \\
& \\
FL8Y\_Object& & 4FGL J1411.5-6133 & & & & & 4FGL J1913.3+1019 \\
& \\
FL8Y\_Otype& & PSR & & & & & PSR \\
& \\
FL8Y\_Dist& & & 0.1 & & & & 0.15 \\
& \\
FL8Y\_SFoverlap& & 0.12 & & & & & 0.06\\
 \hline
 \hline
\end{tabular}}
\label{tab:SNRcand}
\end{table*}

\begin{figure*}[!p]
\centering   
\includegraphics[width=\hsize]{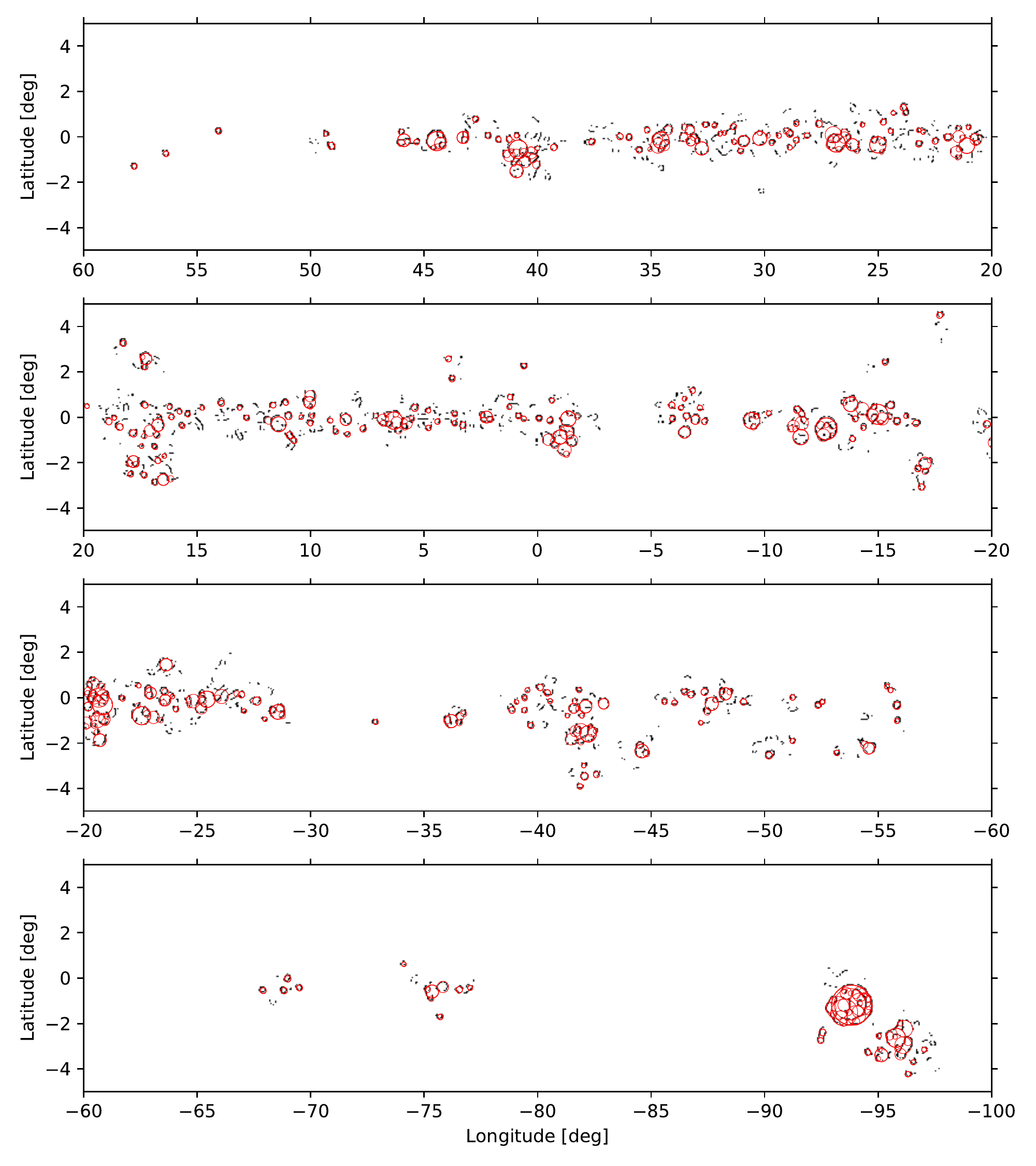}
\caption{Objects detected in the HGPS significance map -- Black contours correspond to the edge mask and red circles to the objects detected in Hough space.}
\label{fig:allsky}
\end{figure*}

\end{document}